\documentclass[envcountsame]{llncs}
\date{\today}

\usepackage[utf8]{inputenc}
\usepackage{amsmath}
\usepackage{amsfonts}
\usepackage{amssymb}
\usepackage{cite}
\usepackage{float}
\usepackage[colorlinks]{hyperref}
\hypersetup{
  colorlinks, citecolor=Blue, linkcolor=Blue, urlcolor=Blue}
\usepackage{cleveref}
\usepackage[dvipsnames]{xcolor}
\usepackage{algorithm}

\usepackage{algpseudocode}
\usepackage{xspace}
\usepackage{multirow}
\usepackage{diagbox}
\usepackage{multirow}
\usepackage{array}
\usepackage{siunitx}
\usepackage{booktabs}
\algnewcommand\algorithmicinput{\textbf{Parameters:}}
\algnewcommand\Params{\item[\algorithmicinput]}
\usepackage{derivative}
\usepackage{mathtools}
\usepackage{bbm}
\usepackage{tikz}
\usepackage{forest}

\newtheorem{claim}[theorem]{Claim}

\let\lncsendproof\endproof
\renewcommand{\endproof}{\qed\lncsendproof}

\newcommand{\polylog}{\mathrm{polylog}}

\newcommand{\opt}{\ensuremath{\mathrm{OPT}}}
\newcommand{\eps}{\varepsilon}

\newcommand{\dist}{\ensuremath{\mathrm{dist}}}

\renewcommand{\tilde}{\widetilde}
\DeclareMathOperator*{\ex}{\mathbb{E}}

\newcommand{\unsat}{\ensuremath{\mathrm{unsat}}}

\newcommand{\transpose}{\intercal}

\newcommand{\cH}{\ensuremath{\mathcal{H}}}

\newcommand{\cI}{\ensuremath{\mathcal{I}}}

\newcommand{\cJ}{\ensuremath{\mathcal{J}}}

\newcommand{\cK}{\ensuremath{\mathcal{K}}}



\title{Min-CSPs on Complete Instances III: Nearly Optimal Characterizations of Approximation for Symmetric CSPs}
\author{Yihwan Chae\inst{1} \and Chenglin Fan\inst{1} \and
Donghyeon Lee\inst{1} \and Euiwoong Lee\inst{2}}
\institute{Seoul National University, Seoul, South Korea\\
\email{benedict0724@snu.ac.kr, fanchenglin@snu.ac.kr, lisifu@snu.ac.kr}
\and
University of Michigan, Ann Arbor, USA\\
\email{euiwoong@umich.edu}}
\date{}
\hypersetup{pdfauthor={},bookmarksdepth=2}

\begin{document}
\maketitle
\begin{abstract}

    This paper studies {\em complete} Min-$k$-CSPs: an instance on $n$ variables has one nontrivial constraint for every $k$-set of variables (so $\binom{n}{k}$ constraints in total), and the goal is to find an assignment that minimizes the number of unsatisfied constraints. They are motivated by the connections to machine learning and data science, as well as the possibility of being testbeds for new algorithmic techniques.
    Previous recent results~\cite{ALS25, ALMS25} presented polynomial-time algorithms for the {\em decision version} of every complete Boolean $k$-CSP, which is a necessary condition for any multiplicative approximations for the minimization version, but nontrivial approximation results have been limited to {\em fragile CSPs}~\cite{KS09}, 2-SAT~\cite{ALS25}, and NAE-3-SAT~\cite{ALMS25}, leaving the approximability of most complete Boolean Min-CSPs open, including complete Min-$k$-SAT. 

    In this paper, we resolve this question, up to nearly matching factors, for the natural class of {\em symmetric} Boolean predicates.  
    Given $A \subseteq \{ 0, \dots, k \}$, Min-CSP($A$) is a Min-$k$-CSP where each constraint is specified by $k$ literals and is satisfied if the number of true literals belongs to $A$. Let $w(A)$ be the maximum number of consecutive integers in $A$.
    Our main results for complete Min-CSP($A$) are (1) a QPTAS for $w(A) \leq 2$ and (2) an $\widetilde{O}(n^{w(A)-3})$-approximation for $w(A) \geq 3$, as well as almost matching hardness results for the latter. Our QPTAS for $w(A)=2$ significantly improves the previous $O(1)$-approximation for Min-2-SAT and the $\polylog(n)$-approximation for Min-NAE-3-SAT, as well as extends the PTAS for fragile CSPs (i.e., $w(A) = 1$).

	\end{abstract}
\setcounter{algorithm}{1}

\section{Introduction}

Constraint Satisfaction Problems (CSPs) provide a unified framework for many of the most central problems in theoretical computer science, including SAT, graph coloring, and basic graph cut problems.  An instance consists of variables taking values from a fixed alphabet and a collection of local constraints, and the goal is either to decide whether all constraints can be simultaneously satisfied or to optimize the number of satisfied constraints.  

The maximization versions, Max-CSPs, have been studied extensively.
On general instances, Raghavendra's result gives SDP-based approximation algorithms, which are optimal assuming the Unique Games Conjecture~\cite{Raghavendra08}.
Beyond general instances, Max-CSPs on structured instances have been actively studied as well. 
In particular, on dense instances, where an $n$-variable $k$-CSP has $\Omega(n^k)$ constraints, every Max-CSP admits a PTAS through various algorithmic techniques such as random sampling~\cite{AKK95,BFdLVK03,ADLVKK03,dlVKKV05,MS08,KS09,BHHS11,Yar14,MM15,FLP16}, regularity lemmas~\cite{FK96,COCF10,OGT13,JST21}, and convex-programming hierarchies~\cite{dlVKM07,AKK+08,BRS11,GS11,YZ14,AJT19,JQST20,BBK+21}.

There is also a large literature on Min-CSPs, whose objective is to minimize the number of violated constraints. While it is the same as Max-CSPs in terms of the exact optimization, their multiplicative approximations behave quite differently: a small additive error that is often harmless for an approximation for Max-CSPs can be prohibitively large for Min-CSP when the optimum value is small (e.g., if the instance is satisfiable, any multiplicative approximation for Min-CSPs must return a satisfying assignment).  For example, Min-2-SAT (also known as Min-2CNF-Deletion) admits an $O(\sqrt{\log n})$-approximation~\cite{ACMM05} and is hard to approximate within any constant factor under the Unique Games Conjecture~\cite{Khot02} on general instances, while the corresponding maximization problem is constant-approximable.  More broadly, the approximability of Boolean Min-CSPs on general instances has a rich structural theory~\cite{KSTW01}, but much less is known for structured instances.

This paper studies Min-CSPs on \emph{complete} instances.  A complete $k$-CSP has exactly one nontrivial constraint on every $k$-set of variables, which is a special case of dense instances. Recent studies have shown that their approximability can behave very differently from dense instances and can be testbeds for new algorithmic ideas.  
Complete instances also capture several natural optimization problems arising in machine learning and data science, including Correlation Clustering~\cite{CALN22,CALLN23,CCAL+24}, Low-Rank Approximation~\cite{BBB+19,CAFG+24}, and various metric fitting problems~\cite{CFLdM25,BCAH+23,anderson2026additive}.

Anand, Lee, and Sharma~\cite{ALS25} initiated the systematic study of complete Min-CSPs.
For the decision version whose goal is to determine whether the instance is completely satisfiable, they proved that every Boolean complete $k$-CSP admits a quasi-polynomial-time algorithm, which was improved to polynomial time by Anand, Lee, Mazzali, and Sharma~\cite{ALMS25}. 

However, the known results for approximations have been limited. Karpinski and Schudy~\cite{KS09} gave a PTAS for {\em fragile} Min-CSPs on more general dense instances; the fragility here means that changing a value of a single variable always makes a satisfied constraint unsatisfied (e.g., $k$-LIN). Beyond fragile CSPs, many natural dense Min-CSPs tend to inherit the hardness of general instances, but few approximation algorithms for complete instances have been designed. 
Anand et al.~\cite{ALS25} gave an $O(1)$-approximation for complete Min-2-SAT, contrasting with an $\omega(1)$-hardness for dense Min-2-SAT, and the subsequent paper \cite{ALMS25} gave a quasi-polynomial-time polylogarithmic approximation for complete Min-NAE-3-SAT, contrasting with the NP-hardness of the decision Min-NAE-3-SAT on dense instances. To the best of our knowledge, these two are the only non-fragile Min-CSPs for which nontrivial approximation algorithms for complete instances have been designed, and both \cite{ALS25} and \cite{ALMS25} left open the approximability of most complete Boolean Min-CSPs, including the basic family of complete Min-$k$-SAT for $k\geq 3$. 

In this paper, we resolve this question, up to nearly matching factors, for the natural class of {\em symmetric} Boolean predicates, where the satisfaction of a constraint only depends on the number of true literals.   
Formally, let $A \subseteq \{ 0, \dots, k \}$. Min-CSP($A$) is a Boolean Min-$k$-CSP where each constraint is specified by $k$ literals (variables or their negations)
and is satisfied if the number of true literals belongs to $A$. This framework captures many standard CSPs, including $k$-SAT ($A=\{1,\ldots,k\}$), NAE-$k$-SAT ($A=\{1,\ldots,k-1\}$), and $k$-LIN ($A=\{0,2,\dots\}$).

Our characterization depends on one simple parameter.  For $A\subseteq\{0,1,\ldots,k\}$, let $w(A)$ be the maximum length of an interval of consecutive integers contained in $A$.  We call $A$ \emph{trivial} if $A=\emptyset$ or $A=\{0,1,\ldots,k\}$ and \emph{nontrivial} otherwise.  
The case $w(A) = 1$ coincides with the fragility, so the work of Karpinski and Schudy~\cite{KS09} implies a PTAS. Our main result is the following theorem, showing that the value of $w(A)$ almost completely determines the approximability of complete Min-CSP($A$).

\begin{theorem}\label{thm:main}
Fix $k\ge 2$ and $A\subseteq\{0,1,\ldots,k\}$.
For a complete Min-CSP($A$) instance $\cI$ on $n$ variables, the following hold. 
\begin{enumerate}
\renewcommand{\labelenumi}{(\alph{enumi})}

\item If $1\le w(A)\le 2$, then for every $\varepsilon\in(0,1)$ there is a deterministic $(1+\eps)$-approximation algorithm running in time
$
        n^{\operatorname{poly}(\log n,1/\varepsilon)}.
$

\item If $3\le w(A)\le k$, then there is a randomized polynomial-time 
$O(n^{w(A)-3}\log^{9/2} n)$-approximation algorithm that succeeds with probability at least $2/3$. 

\item If $A$ is nontrivial and $w(A)=3$, then complete Min-CSP($A$) is NP-hard to approximate within any constant factor $\rho>1$, assuming the Unique Games Conjecture.
\item If $A$ is nontrivial and $w(A)\ge4$, then for every fixed $\varepsilon\in(0,w(A)-3)$, complete Min-CSP($A$) is NP-hard to approximate within factor $n^{w(A)-3-\varepsilon}$.
\end{enumerate}
\end{theorem}

A recently announced proof of the Unique Games Conjecture by \mbox{OpenAI}~\cite{OpenAI26UGC} would make part~(c) unconditional.
\begin{table}[H]
\caption{Approximation landscape for complete Min-CSP($A$). Our new results in boldface. }
\label{tab:width-classification}
\centering
\small
\begin{tabular}{@{}>{\raggedright\arraybackslash}p{0.12\textwidth}|>{\raggedright\arraybackslash}p{0.41\textwidth}|>{\raggedright\arraybackslash}p{0.38\textwidth}@{}}
\toprule
\multicolumn{1}{@{}c|}{Width} & \multicolumn{1}{c|}{Algorithm} & \multicolumn{1}{c@{}}{Hardness} \\
\midrule
$w(A)=1$ & PTAS~\cite{KS09} & Exact optimization NP-hard~\cite{ALS25} \\
\midrule
$w(A)=2$ & {\bf QPTAS} (Theorem~\ref{thm:qptas}) & Exact optimization NP-hard~\cite{ALS25} \\
\midrule
$w(A)=3$ & {\bf $O(\log^{9/2} n)$-approx.}  (Theorem~\ref{thm:wge3}) &
{\bf $\omega(1)$-UG-hard} (Lemma~\ref{lem:binary-transfer-h3}) \\
\midrule
$w(A)\ge4$ & {\bf $O(n^{w(A)-3}\log^{9/2} n)$-approx.} (Theorem~\ref{thm:wge3}) & {\bf $(n^{w(A)-3-\varepsilon})$-NP-hard} (Lemma~\ref{lem:lower-poly}) \\
\bottomrule
\end{tabular}
\end{table}

Our QPTAS for $w(A) = 2$ significantly improves (polynomial-time) $O(1)$-approximation for Min-2-SAT and (quasi-polynomial-time) $\polylog(n)$-approximation for NAE-3-SAT~\cite{ALS25, ALMS25}. Furthermore, our upper and lower bounds almost match for every width. For $w(A) = 2$, perhaps the only significant open problem left is improving the QPTAS to a PTAS. 
For $w(A) = 3$, the optimal approximation ratio lies between $\omega(1)$ and $\polylog(n)$. 
For $w(A) \geq 4$, the optimal approximation ratio is $n^{w(A)-3}$ up to a factor of $n^{\eps}$ for any $\eps > 0$.

    Our algorithmic framework proceeds in two phases, where the first phase computes (a list containing) an assignment close to the optimal one in Hamming distance, and the second phase uses it for a good approximation for the CSP. Such two-phase frameworks were implicit in previous work for fragile Min-CSPs~\cite{KS09, meot2023voting} and related problems of low-rank approximation and metric fitting~\cite{BBB+19, CAFG+24}, where both phases primarily use the {\em voting/greedy process} effective only for fragile CSPs or similar constraint structures. We introduce broader techniques in both phases to handle general constraint structures, including direct product testing~\cite{IJKW10} and polynomial optimization over truncated simplices~\cite {Barman18}. We believe that our framework and techniques are of independent interest and may help unify and advance the understanding of complete Min-CSPs beyond the Boolean alphabet. 

\paragraph{Concurrent work.}
In independent and concurrent work, Cai, Jiang, and Liu~\cite{CJL26} study complete Boolean Min-$k$-CSPs in which the predicate can vary from one $k$-set of variables to another.
The accepting-subcube dimension of a predicate is the largest number of inputs that can be left free while fixing the others to force acceptance.
When every predicate has dimension at most $h$, their randomized polynomial-time approximation ratios are
\[
\begin{cases}
 O(1), & h=1,\\
 O\!\left(\sqrt{\log\!\left(2+\opt/\binom n{k-3}\right)}\right), & h=2,\\
 O((n/\log(n+2))^{h-2}), & h\ge3,
\end{cases}
\]
where $\opt$ is the minimum number of violated constraints.

A nontrivial symmetric predicate specified by $A$ has accepting-subcube dimension $w(A)-1$, even after literal negations.
Both approaches look for assignments close to an optimum by fixing a set of variables and working with the constraints that contain that set.
Cai, Jiang, and Liu then extend low-cost assignments on a random subset by predicting each remaining bit separately, while our randomized algorithm combines assignments from many subsets using direct product testing.
In one of their repair routines, a semidefinite relaxation determines which variables enter a weighted Min-2-SAT instance.
Our repair for $w(A)\ge3$ builds the weighted instance on all variables and adds a penalty for each flip.

\section{Technical overview}\label{sec:overview}

In this section, we give a more technical overview of the proof, using complete Min-3-SAT and Min-2-SAT as primary examples. At the highest level, our algorithmic framework proceeds in two phases. In the first phase, we compute a {\em close-list}, which is a list of assignments that contains an assignment close to the optimal assignment in Hamming distance (e.g., $\widetilde{O}(\opt^{1/k})$ for $k$-SAT). And the second phase uses such a close assignment to compute another assignment that violates a small number of constraints. 

The idea of such a two-phase framework is not entirely new --- it is implicit in previous results for fragile Min-CSPs~\cite{KS09, meot2023voting} and related problems of low-rank approximation and metric fitting~\cite{BBB+19, CAFG+24}. However, to the best of our knowledge, the clear demarcation and relationship between the two phases were never made explicit, partly due to the fact that both phases use the same algorithmic idea of the {\em voting/greedy process}, which is effective only for fragile CSPs or similar constraint structures. 
One of our conceptual contributions is a clear separation between the two phases, which allows us to use {\em diverse tools for different phases and situations}. 

For instance, our (randomized) polynomial-time algorithm to compute the close-list in the first phase uses the celebrated {\em direct product testing}~\cite{IJKW10} primarily used in complexity theory, making it one of the rare applications of direct product testing to approximation algorithms. 
\Cref{sec:overview-close} further explains this part, which also complements our hardness reduction in \Cref{sec:overview-hardness}.

For the second phase, starting from an assignment close to the optimum in Hamming distance, our algorithms use different tools for two different cases.
For complete Min-3-SAT, we give a technical reduction to a general instance of Min-2-SAT for a $\polylog(n)$-approximation, exploiting the bounded Hamming distance $H$ to the optimal solution and the fact that an additive error of $O(H^3)$ is tolerable as $H = \widetilde{O}(\opt^{1/3})$.
For complete Min-2-SAT, the bounded $H$ allows us to combine classical tools in the area, such as an additive QPTAS for quadratic polynomial optimization over the truncated simplex~\cite{Barman18}. These two algorithms are presented in \Cref{sec:overview-3sat} and \Cref{sec:overview-2sat}, respectively. 

Finally, in \Cref{sec:overview-pivot}, we show how to generalize our ideas for 3-SAT and 2-SAT to general Min-CSP($A$) with $w(A) \geq 3$ and $w(A) \leq 2$ respectively, based on the simple {\em pivot} idea.

\subsection{Hardness}
\label{sec:overview-hardness}
To motivate the notion of close-lists, let us first describe the hardness for complete Min-3-SAT, ruling out an $O(1)$-approximation assuming the Unique Games Conjecture (UGC). The source problem is a general (not necessarily complete) Min-2-SAT, which is $\omega(1)$-hard to approximate assuming the UGC~\cite{Khot02}. 

Given a general Min-2-SAT instance $\cJ$ on variables $X$ and constraints $E$, with at most one clause per variable pair, we transform it into a complete Min-3-SAT instance $\cI$. First, add a set $Y$ of auxiliary variables with $|Y| = Q$ for a large enough $Q$ to be determined. 
For every $y\in Y$ and an original 2-SAT constraint on $\{ u, v \} \subseteq X$, add the 3-SAT clause on $\{y,u,v\}$, which is violated if and only if the 2-SAT constraint on $\{ u, v \}$ is violated and $y = 0$. (E.g., given $(\neg u \vee v)$, create $(y \vee \neg u \vee v)$. Throughout the paper, we will use $1$ as True and $0$ as False.) Call them {\em real clauses}. Thus, when $y=0$, this 3-SAT clause is violated exactly if the source 2-SAT constraint is violated. 
For all other triples of variables in $X \cup Y$, add the 3-SAT clause where every variable is negated. (Indeed, literals for the original variables do not matter.) Call them {\em dummy clauses}.

For completeness, if there is an assignment $x \in \{0, 1\}^X$ that violates $\beta$ 2-SAT constraints, extending it to $\cI$ by simply setting every auxiliary variable $0$, violates at most $Q\beta + |X|^3$ clauses; the number of violated real constraints is exactly $Q \beta$, and a dummy constraint is violated only if it is entirely contained in $X$. So we will take $Q$ large enough so that the additive $|X|^3$ becomes small compared to $Q \beta$. 

For soundness, the fact that every dummy constraint has its auxiliary variables negated forces that any good assignment for the 3-SAT instance $\cI$ must have almost all auxiliary variables set to $0$. Then, any good assignment to $\cI$ will imply a good assignment to $\cJ$ by using real constraints. 

So, in order to rule out a $t$-approximation for a (large) constant $t$, it suffices to have $Q$ such that $Q\beta = \Theta(|X|^3)$, which implies $\opt = \Theta(|X|^3)$. In this regime, the hard instance from complete Min-3-SAT completely reveals the assignment for auxiliary variables (i.e., every $y \in Y$ is assigned $0$ in the optimal solution), and the complexity of this instance is entirely hidden in the small set of variables $X$ with $|X| = \Theta(\opt^{1/3})$. Is it an artifact of the reduction, or is there a way to reverse-engineer this observation and use it algorithmically?

\subsection{Close-lists}
\label{sec:overview-close}
Somewhat surprisingly (at least to us), we show that any instance of complete Min-CSP can be reduced to one similar to the hard instance above. More specifically, for any complete Min-CSP, we present efficient algorithms that construct a {\em close-list}, which contains an assignment whose Hamming distance to the optimal assignment is small. For Min-3-SAT, the guaranteed Hamming distance is $\widetilde{O}(\opt^{1/3})$, nearly matching the regime of the hard instance discussed above up to polylog factors. (Throughout this paper $\widetilde{O}(.)$ hides $\polylog(n)$ factors.)  

We give two algorithms for computing close-lists, a deterministic quasi-polynomial-time algorithm and a randomized polynomial-time algorithm. The deterministic quasi-polynomial-time algorithm extends the combinatorial guessing and branching framework of \cite{ALS25}, which worked only for satisfiable instances, to unsatisfiable instances by carefully accounting for how violated constraints in the optimal solution affect the framework.

In this overview, we focus on the randomized polynomial-time algorithm, which uses the celebrated direct-product testing~\cite{IJKW10} together with the VC-dimension bound on the space of satisfying solutions for complete Min-CSPs~\cite{ALS25}.

In our context, the goal of direct product testing is to recover a hidden string $x \in \{ 0, 1\}^n$ from noisy observations of the form $x|_W$ for $W \subseteq [n]$ (i.e., the restriction of $x$ on $W$). 
Impagliazzo, Jaiswal, Kabanets, and Wigderson~\cite{IJKW10} showed that, 
given oracle access to table $Y$ that maps each $W \in \binom{[n]}{\ell}$ to $Y(W)\in\{0,1\}^W$ with the guarantee
\[
    \Pr_{W\in\binom {[n]}\ell}[Y(W)=x|_W] \geq \eps,
\]
there is a randomized polynomial-time algorithm that outputs, with probability $\Omega(\varepsilon)$, an assignment $z\in\{0,1\}^n$ satisfying
\[
        \dist_H(z,x)\le
        O\left(\frac{n\log(1/\varepsilon)}{\ell} \right),
\]
where $\dist_H$ denotes the Hamming distance. 
Given a budget $B$, we show how to use this result to compute a list of polynomially many assignments so that, for any assignment $x$ for complete Min-3-SAT on $n$ variables violating at most $B$ constraints, the list contains an assignment whose Hamming distance to $x$ is $\tilde{O}(B^{1/3})$. 

Let $\ell  = \Theta(\frac{n}{B^{1/3}})$.  
For a uniformly random $\ell$-set $W$, the expected number of violated clauses of $x$ induced inside $W$ is at most
\[
        B\left(\frac{\ell}{n}\right)^3=O(1),
\]
so with constant probability the restriction $x|_W$ satisfies the induced complete instance on $W$. Since the VC-dimension of the space of satisfying assignments for any complete $r$-CSP is at most $r-1$~\cite{ALS25}, a satisfiable complete Boolean $3$-CSP on $W$ has only $O(\ell^2)$ satisfying assignments, which can be enumerated in polynomial time~\cite[Section~5]{ALMS25}.  Therefore, if for each $W$ we sample one satisfying assignment of the induced instance when one exists (and output an arbitrary string otherwise), it gives (randomized) $Y$ such that $Y(W)$ agrees with $x|_W$ on at least an $\eps = \Omega(1/\ell^2)$ fraction of all $W \in \binom{[n]}{\ell}$.

Therefore, the above direct-product decoder recovers, in polynomial time, an assignment $z$ satisfying
\[
        \dist_H(z,x)\le O\left(\frac{n\log \ell}{\ell}\right)
        =O(B^{1/3}\log n),
\]
with probability $\Omega(\eps)$, and repeating it polynomially many times ensures that an assignment close to $x$ is included in the list with high probability. 

\subsection{Complete Min-3-SAT}
\label{sec:overview-3sat}

We next explain our algorithm for complete Min-3-SAT on the variable set $V$, which is a special case of $w(A)=3$ of \Cref{sec:w-ge-3}.  Let $x^*$ be an optimum assignment and use the close-list algorithm to obtain a candidate reference assignment $z$ with
\[
        R^*:=\{v:z_v\ne x^*_v\},
        \qquad |R^*| = H,
        \qquad H=O(\opt^{1/3}\log n).
\]
(Let us assume we know $H$ and $\opt$ by guessing, though $R^*$ is not known.)
The goal is now to find a low-cost correction $s\in\{0,1\}^V$, where $s_v=1$ means flipping $z_v$.

Our algorithm proceeds via reducing it to weighted Min-2-SAT, which admits a polylogarithmic approximation~\cite{ACMM05}.  
A violated clause of $z\oplus s$ can involve zero, one, two, or three flipped variables. 
A starting observation for this reduction is that the two extreme cases can be dealt with easily as follows.
\begin{itemize}
    \item Clauses involving three flipped variables: if a correction flips $L$ variables, there are only $O(L^3)$ clauses involving three flipped variables, so as long as we ensure $L = \widetilde{O}(H)$, the number of such clauses will be $\widetilde{O}(\opt)$.  
    
    \item Clauses involving zero flipped variables: in the optimal correction $s^*$, every violated clause with zero flipped variables is violated in $x^*$ as well, so the number of such clauses is at most $\opt$. (Our formal analysis for general $s$ considers the union of the flips by $s$ and $s^*$.)
\end{itemize}
The main task is therefore to encode the effect of one or two flipped variables by a weighted Min-2-SAT instance on the correction variables $s_v$.
The reduction has the following three kinds of terms. They are somewhat technical but naturally justified; note that the goal is to design them so that (1) the value for optimal $s^*$ is upper-bounded and (2) the value for general $s$ is lower-bounded with respect to the number of clauses they violate. 

\begin{enumerate}
    \item First, add a unary penalty of weight $H^2$ for setting $s_v=1$, ensuring that any polylogarithmic approximate solution for weighted Min-2-SAT flips only $\widetilde{O}(H)$ variables.
    
    \item Second, in order to account for two-flip violations, for each pair $P=\{u,v\}$ and each local correction pattern $\tau\in\{0,1\}^P$, count the number of third variables $w$ such that the 3-SAT clause on $P\cup\{w\}$ would be violated if $s|_P=\tau$ and $s_w=0$.  
    
    It is natural to use this number as the weight for the 2-SAT constraint for $P$ forbidding $s|_P = \tau$ (e.g., $(u \vee v)$ if $\tau = (0, 0)$), but it can potentially make the 2-SAT value of the optimal solution $s^*$ very high, because if $w \in R^*$ (i.e., $s^*_w = 1$), it is possible that $w$ counts for $s^*|_P$ even when $s^*$ satisfies the constraint for $\{ u, v, w \}$. In order to prevent this over-counting, the actual weight for the 2-SAT constraint corresponding to $\{ u, v \}$ and $\tau$ is the number of contributing $w$'s minus the slack $H$ (or $0$ if it is negative). 
    
    This subtraction, in turn, may hurt lower-bounding the cost of general $s$, but its effect can be controlled as follows. Consider a pair $\{ u, v \}$ flipped in $s$. If the number of $w$'s counted towards flipping $u, v$ is at least $2H$ (say $\{ u, v \}$ is {\em large} in this case and {\em small} otherwise), then the weight after subtraction is at least half the original count, so the subtraction only loses a multiplicative factor of $2$. Otherwise, the total number of violated clauses involving $u, v$, and an unflipped variable is at most $2H$, so by summing over all flipped pairs $\{u, v \}$, the total number of violated clauses containing a small flipped pair is $O(HL^2)=\widetilde{O}(H^3)$, so we incur an additive error of $\widetilde{O}(\opt)$.

    \item Third, for one-flip violations, fix a variable $v$ and value $a$, and consider the graph $\cH_v^a$ on the other variables whose edges $\{u,w\}$ are those for which the clause on $\{v,u,w\}$ is violated by $s_v=a$ and $s_u=s_w=0$. 
    
    Then the natural 2-SAT constraint would be forbidding $v \leftarrow a$ with weight $|E(\cH_v^a)|$. Again, this potentially makes the value for the optimal solution $s^*$ very high, because even a satisfied constraint on $\{ u, v, w \}$ with one or two variables flipped could be counted. (E.g., if only $w$ is flipped, then this constraint is counted towards the weight of $u$ and the weight of $v$.)
    In order to mitigate this, we introduce a slightly more technical subtraction scheme: a vertex $u \in V(\cH_v^a)$ is {\em heavy} if its degree is larger than $4H$ and {\em light} otherwise, and the weight of the forbidding $v \leftarrow a$ is 
    \[
    \max \{ | \{ (u, w) \in E(\cH_v^a) : \mbox{both } u, w \mbox{ are light} \} | - 4H^2, 0 \}.
    \]
    A more involved case analysis ensures that this weight achieves the goal of upper-bounding the 2-SAT value of $s^*$ while lower-bounding the 2-SAT value of $s$ with respect to their actual violations in the 3-SAT instance, up to polylogarithmic multiplicative and $\widetilde{O}(H^3)$ additive errors. 
\end{enumerate}
Therefore, using the known $\polylog(n)$-approximation for weighted Min-2-SAT gives a correction $s$ whose 2-SAT value is at most $\widetilde{O}(\opt + H^3)$, and in the complete 3-SAT instance, it violates at most 
$\widetilde{O}(\opt + H^3) = \widetilde{O}(\opt)$ constraints.

\subsection{Complete Min-2-SAT}
\label{sec:overview-2sat}

For complete Min-2-SAT, one can obtain a stronger result by appealing to tools from polynomial optimization. Let $x^*$ be optimum and let $z$ be a close-list candidate with
\[
        R^*:=\{v:z_v\ne x^*_v\},
        \qquad |R^*| = H,
        \qquad H=O(\opt^{1/2}\log n).
\]
(Again, without loss of generality, $H$ is known but $R^*$ is not.)
Let $\unsat_{\cI}(x)$ be the number of violated constraints of instance $\cI$ for assignment $x$.
For a correction vector $s \in \{0, 1\}^V$ as before, where $s_v = 1$ denotes flipping $z_v$ to $1 - z_v$, the objective value given $z$ and $s$ can be expressed exactly with a quadratic set function:
\[
        \unsat_{\cI}(z\oplus s)
        =\unsat_{\cI}(z)
        +\sum_{i}a_i s_i+
        \sum_{i, j} b_{ij} s_i s_j .
\]
The linear coefficients $a_i$'s can be of order $\Omega(n)$, but because the instance is complete and has only one constraint on each pair, the pair coefficients satisfy $|b_{ij}|=O(1)$.

Thus the problem becomes: minimize a quadratic function
\[
        F(s)=\sum_{i}a_i s_i+
        \sum_{i,j}b_{ij} s_i s_j
\]
over the sliced Boolean hypercube $\{ s \in \{ 0, 1 \}^V : \| s \|_1 = H \}$, up to additive error $\eta H^2$. If we let $p = s / H$, we have $F(s) = H^2 f(p)$ where  
\[
f(p) = \sum_{i} (a_{i}/H) p_i + \sum_{i,j} b_{ij} p_ip_j, 
\]
so minimizing $F(s)$ over the sliced Boolean hypercube becomes essentially equivalent to minimizing $f(p)$ in the truncated simplex
\[
        \Delta_H=\left\{p\ge0:\sum_i p_i=1,\ p_i\le\frac1H\right\}.
\]
The result of Barman~\cite{Barman18}, based on an approximate Carathéodory's theorem, shows that an additive $\eta$-approximation for $f$ (so an additive $\eta H^2$-approximation for $F$) can be found in time $n^{O(\log n/\eta^2)}$. The basic idea is: 
letting $C$ be the matrix corresponding to the quadratic part of $f$ (i.e., $\sum_{i,j} b_{ij}p_ip_j = p^T C p$) and $p^*$ be an optimal solution, one can guess $v$ with $\| v - Cp^* \|_{\infty} \leq O(\eta)$ by trying all convex combinations of $O(\log n/\eta^2)$ columns of $C$, which essentially reduces the problem to a linear program over $\Delta_H$. Therefore, using $H = O(\opt^{1/2} \log n)$ and letting $\eta = 1 / \polylog(n)$, one can bring down the additive error small enough compared to $\opt$, giving a QPTAS. The algorithm of Lipton, Markakis, and Mehta~\cite{LMM03} for approximate Nash equilibria uses a similar enumeration of small multisets.

It is also instructive to see why this method cannot yield a QPTAS for Min-3-SAT. One can similarly model Min-3-SAT as minimizing a cubic polynomial
\[
        F(s)=\sum_{i}a_i s_i+
        \sum_{i,j}b_{ij} s_i s_j + 
        \sum_{i,j,k}c_{ijk} s_i s_j s_k,
\]
over the sliced Boolean hypercube up to an additive error of $\eta H^3$, which is equivalent to minimizing 
\[
        f(p)=\sum_{i}(a_i/H^2) p_i+
        \sum_{i, j}(b_{ij}/H) p_ip_j+
        \sum_{i,j,k}c_{ijk} p_ip_jp_k,
\]
over the truncated simplex up to an additive error $\eta$. The crucial difference is that while $|c_{ijk}| \leq O(1)$, the normalized quadratic coefficient $|b_{ij}/H|$ can be $\Omega(n/H)$, which is not bounded by $O(1)$. While the polynomial optimization framework can handle large linear coefficients, unbounded quadratic coefficients make the use of this framework infeasible.

\subsection{Width classification}
\label{sec:overview-pivot}

We finally explain how the preceding special cases of 2-SAT and 3-SAT extend to an arbitrary symmetric predicate $A\subseteq\{0,1,\ldots,k\}$.  Put
\[
        r:=\max\{2,w(A)\},
        \qquad q:=k-r .
\]
The pivot step in \Cref{sec:pivot} reduces the effective arity from $k$ to $r$.  Fix a target assignment $x$ of cost at most $T$.  For a $q$-set $P$, let $d_P(x)$ be the number of constraints violated by $x$ whose $k$-variable set contains $P$.  Double counting gives
\[
        \sum_{P\in\binom Vq} d_P(x)
        =\binom kq\unsat_{\cI}(x)
        \le \binom kq T,
\]
so some pivot $P$ satisfies $d_P(x)\le O(T/n^q)$.  The algorithms try all $P$ and all assignments on $P$, which yields an induced complete $r$-CSP instance on the remaining $n-q$ variables. 
It is simple to check that every induced constraint is nontrivial; if one of the induced constraints accepted every assignment on its $r$ variables, then $A$ would contain an interval of length $r+1$, contradicting $r\ge w(A)$.  Hence every induced constraint forbids at least one local assignment, and for a correct guess of $P$ and its assignment, the close-list machinery applies with residual budget
\[
        B=O\left(\max\left\{1,\frac{T}{n^q}\right\}\right).
\]
It returns a reference assignment within
\[
        H=O\left(\left(\max\left\{1,\frac{T}{n^q}\right\}\right)^{1/r}\log n\right)
\]
of the target assignment $x$.  

Given an assignment $z$ close to $x$ in Hamming distance, our algorithms for $w(A) \geq 3$ and $w(A) \leq 2$ naturally extend the aforementioned algorithms for 3-SAT and 2-SAT respectively. 
For $w(A)\geq 3$, the algorithm proceeds via a reduction to weighted Min-2-SAT, essentially ignoring constraints involving at least three flipped variables. Thus the straightforward extension of the 3-SAT analysis gives a solution of cost
\[
        \widetilde{O}(n^{k-3}H^3),
\]
which is an $\widetilde{O}(n^{w(A)-3})$-approximation, because, for $T = \Omega(n^{k-w(A)})$, 
\[
\begin{aligned}
        n^{k-3}H^3
        &\le \widetilde{O}\left(
        n^{k-3}\left(\frac{T}{n^{k-w(A)}}\right)^{3/w(A)}
        \right) \\
        &= \widetilde{O}\left(
        n^{w(A)-3}T
        \left(\frac{n^{k-w(A)}}{T}\right)^{(w(A)-3)/w(A)}
        \right) \\
        &= \widetilde{O}\left(
        n^{w(A)-3}T
        \right).
\end{aligned}
\]
The case $T = o(n^{k-w(A)})$ can be solved exactly because one of the induced instances is perfectly satisfiable.

A QPTAS for $w(A) \leq 2$ also proceeds similarly to Min-2-SAT, though there are additional subroutines to handle possible higher-degree terms in the polynomial $F$ and $f$.

\label{ipco:main-end}

\clearpage
\begin{credits}
\subsubsection{\ackname}
We thank the anonymous SODA reviewers for their helpful comments.

GPT 5.5 Pro helped the authors with the following results: (1) The randomized polynomial-time algorithm for computing a close-list, prompted with the authors' two-phase framework and preliminary algorithms for computing the list. (2) The QPTAS for the $w(A)\le 2$ case, using the authors' close-list construction after exposing a pivot and quadratic-minimization formulation. (3) Generalizing the algorithmic and hardness results from 3-SAT and NAE-3-SAT to symmetric CSPs. It was used in generating and revising proofs with human guidance as well. The authors assume responsibility for all content.
\end{credits}
\bibliographystyle{splncs04}
\bibliography{refs}

\clearpage
\appendix
\renewcommand*{\theHsection}{\thesection}
\section{Preliminaries}\label{sec:notation}

Let $V$ be a set of $n$ Boolean variables. We identify $1$ with True and $0$ with False. Throughout the paper, we allow {\em negations}: on each occurrence of a variable $i$, a constraint may use either the literal $x_i$ or the literal $1-x_i$.  
Fix $k \geq 2$ and $A\subseteq\{0,1,\ldots,k\}$. An $A$-constraint is specified by a $k$-set $S\in\binom{V}{k}$ (called the {\em scope}) and for every $i\in S$, a bit $\nu_{S,i}\in\{0,1\}$ indicating whether that occurrence is negated.
It is satisfied when
\[
        \sum_{i\in S} (x_i \oplus \nu_{S,i}) \in A.
\]
A complete Min-CSP($A$) instance $\cI$ has exactly one $A$-constraint on every $k$-set $S\in\binom{V}{k}$.
Thus there are no parallel constraints and no constraint repeats a variable.
In general, a complete $k$-CSP means that every $k$-set receives exactly one nontrivial constraint.
For an assignment $x\in\{0,1\}^V$, let
\[
        \unsat_{\cI}(x)=\#\{S\in\binom{V}{k}: S \text{ is violated by }x\},
        \qquad
        \opt(\cI)=\min_x \unsat_{\cI}(x).
\]
The objective is the unnormalized violated-constraint count.
Dividing all costs by $\binom nk$ gives the equivalent normalized objective.
For two assignments $x,z\in\{0,1\}^V$, let
\[
        \dist_H(x,z):=|\{v\in V:x_v\ne z_v\}|
\]
denote their Hamming distance.

\begin{definition}[width]
For $A\subseteq\{0,1,\ldots,k\}$, define
\[
        w(A)=\max\{\ell:\exists a\text{ such that }\{a,a+1,\ldots,a+
        \ell-1\}\subseteq A\}.
\]
Set $w(\emptyset)=0$.
If $A=\{0,1,\ldots,k\}$, then $w(A)=k+1$.
\end{definition}

Throughout the paper, $k$ and $A$ are fixed.
Constants hidden in asymptotic notation may depend on them, but never on $n$ or on the input instance.
We rely on the following lemma, which shows that the number of satisfying assignments of any complete (not necessarily symmetric) CSP is polynomial, as long as every constraint is violated by some assignment.

\begin{lemma}[Lemma 3.1 of \cite{ALS25}]\label{lem:vc}
Let $\cI$ be a complete, Boolean $k$-CSP on $n$ variables, where every $k$-constraint forbids at least one local assignment.
Then the set of satisfying assignments has size at most
\[
        s_k(n):=\sum_{i=0}^{k-1}\binom ni=O(n^{k-1}).
\]
For fixed $k$, the satisfying assignments can be enumerated in polynomial time.
\end{lemma}

\section{Close-lists for low-cost assignments}\label{sec:close-lists}

In this section, we present algorithms that compute a {\em close-list} --- a set of assignments that includes one close to the hidden optimal assignment in Hamming distance. 
In particular, given a complete Min-CSP($A$) instance of arity $k$, let $r := \max(2, w(A))$ and $q = k - r$. The main theorem of this section, \Cref{thm:pivot} in \Cref{sec:pivot}, presents algorithms that compute a list that contains an assignment $z$ whose Hamming distance from an optimal assignment $x^*$ is at most $O((\opt(\mathcal{I}) / n^q)^{1/r} \log n)$ when $\opt(\cI) = \Omega(n^q)$. 

Key subroutines for this main result are algorithms that, given a complete $r$-CSP instance $\cI$, compute a list that contains an assignment at distance at most $O(\opt(\cI)^{1/r} \log n)$. We present two algorithms, a deterministic quasi-polynomial-time one in \Cref{sec:close-det} and a randomized polynomial-time one in \Cref{sec:close-rand}. Then, \Cref{thm:pivot} easily follows by guessing a {\em pivot} $P$ of $q$ variables as well as $x^*|_P$, and applying them to the complete $(k-q)$-CSP instance $\cI$ on the variable set $V \setminus P$, where each constraint for $W \subseteq V \setminus P$ is obtained from the original constraint for $P \cup W$. 

Finally, in \Cref{sec:smallopt}, we show that the case $\opt(\cI)<\tau n^q$ for a sufficiently small constant $\tau>0$ can be solved exactly in polynomial time.

\subsection{Deterministic close-lists}
\label{sec:close-det}

In this subsection, we present and analyze our deterministic algorithm, proving the following theorem. 

\begin{theorem}
\label{thm:oneforbidden-list}
For every fixed $r\ge2$, given a complete $r$-CSP $\cI$, in which every constraint forbids at least one local assignment, on a variable set $U_0$ of size $m$ and a budget $B\ge1$, there is a deterministic $m^{O(\log m)}$-time algorithm that outputs a list $\mathcal L\subseteq\{0,1\}^{U_0}$ of size $m^{O(\log m)}$.
For every assignment $x$ with $\unsat_{\cI}(x)\le B$, the list contains $z\in\mathcal L$ satisfying
\[
        \dist_H(z,x)\le O(B^{1/r}\log(m+2)).
\]
\end{theorem}

\begin{proof}
Call a complete Boolean $s$-CSP \emph{one-forbidden} if, for every $s$-set $S$, its constraint forbids exactly one local assignment on $S$.
For each $R \subseteq U_0$ with $|R| = r$, choose one forbidden local assignment $b_R \in \{0,1\}^R$; if there is more than one, choose one arbitrarily.
Let $\cK_0$ be the complete one-forbidden $r$-CSP on $U_0$ that forbids exactly $b_R$ on each $R$.
Since every assignment violating a constraint of $\cK_0$ also violates the corresponding constraint of $\cI$, every $x$ with $\unsat_{\cI}(x)\le B$ also satisfies $\unsat_{\cK_0}(x)\le B$.
Thus it suffices to prove the theorem for complete one-forbidden CSPs.
We prove this by induction on $r$.

\paragraph{Algorithm.}
The recursive procedure is shown in Algorithm~\ref{alg:det-close-list}.
The top-level call is \\$\mathrm{CloseList}(r,\cK_0,U_0,B)$.
At a recursive call, $U$ is the current unassigned set; assignments fixed on $U_0\setminus U$ are appended when a branch returns.
For $U'\subseteq U$, $\cK|_{U'}$ denotes the induced subinstance of $\cK$ on $U'$.
In the algorithm, $\beta_S$ denotes the unique forbidden local assignment on an $s$-set $S$ in the current complete one-forbidden $s$-CSP.
For $P\in\binom{U}{s-1}$ and a local assignment $\tau\in\{0,1\}^P$, let
\[
        M_P^\tau:=\{w\in U\setminus P:\beta_{P\cup\{w\}}|_P=\tau\}.
\]
When $P=\{u\}$, we write $M_u^a$ for $M_{\{u\}}^\tau$, where $\tau$ is the local assignment on $\{u\}$ with $\tau(u)=a$.

\begin{algorithm}[H]
\small
\caption{Deterministic recursive close-list procedure}\label{alg:det-close-list}
\begin{algorithmic}[1]
\Require arity $s\ge2$, complete one-forbidden $s$-CSP $\cK$ on $U$, budget $B$
\Procedure{CloseList}{$s,\cK,U,B$}
  \If{$|U| \le 2s$}
    \State \Return all $2^{|U|}$ assignments on $U$ \Comment{base case}
  \EndIf
  \State $\theta\gets \lceil B^{1/s}\rceil$ and $\mathcal L\gets\emptyset$
  \For{each $P\in \binom{U}{s-1}$ and $\tau\in\{0,1\}^P$
      such that $|M_P^\tau|\ge |U|/2^s$}
    \State Define a partial assignment $\alpha$ on $P\cup M_P^\tau$ by
    \[
        \alpha(u)=\tau(u) \quad (u\in P),
        \qquad
        \alpha(w)=1-\beta_{P\cup\{w\}}(w) \quad (w\in M_P^\tau).
    \]
    \State $U'\gets U\setminus (P\cup M_P^\tau)$.
    \State $\mathcal L'\gets \mathrm{CloseList}(s,\cK|_{U'},U',B)$.
    \State Extend each assignment in $\mathcal L'$ by $\alpha$ and add it to $\mathcal L$.
  \EndFor
  \If{$s=2$}
    \State For each $u\in U$, choose $a_u\in\operatorname*{arg\,max}_{a\in\{0,1\}}|M_u^a|$, breaking ties arbitrarily
    \State Set $z_u\gets1-a_u$ for all $u\in U$, and add $z$ to $\mathcal L$
  \Else
    \State For each $P\in\binom U{s-1}$, choose $\tau_P^*\in\operatorname*{arg\,max}_{\tau\in\{0,1\}^P}|M_P^\tau|$
    \State Let $\cK'$ be the complete $(s-1)$-CSP on $U$ forbidding $\tau_P^*$ on each $P$
    \State $B'\gets\lceil sB/\theta\rceil$
    \State add every assignment in $\mathrm{CloseList}(s-1,\cK',U,B')$ to $\mathcal L$
  \EndIf
  \State \Return $\mathcal L$
\EndProcedure
\end{algorithmic}
\end{algorithm}
The algorithm has two types of steps.
\begin{itemize}
\item The branching step guesses an assignment $\tau$ on $P$.
If the target assignment really equals $\tau$ on $P$, then every $w\in M_P^\tau$ must take the value $1-\beta_{P\cup\{w\}}(w)$ unless the constraint on $P\cup\{w\}$ is violated.
\item The closing step handles the case where no such good branch exists.
For $s=2$, the algorithm adds the majority-type assignment $z$.
For $s\ge3$, it forms the lower-arity one-forbidden instance $\cK'$ from the most frequent assignments $\tau_P^*$ and calls the same procedure with arity $s-1$.
\end{itemize}

\paragraph{Analysis.}
Fix an assignment $x\in\{0,1\}^{U_0}$ with $\unsat_{\cI}(x)\le B$.
Then $\unsat_{\cK_0}(x)\le B$.
We define a correct same-arity path in the recursion tree of
$\mathrm{CloseList}(r,\cK_0,U_0,B)$ with respect to this fixed assignment $x$.
A node on this path is described by its current unassigned set $U$ and by the partial assignment $\rho$ already fixed on $U_0\setminus U$.
Initially $U=U_0$ and $\rho$ is empty.
At every same-arity node, the current one-forbidden CSP is the restriction of $\cK_0$ to $U$, so $x|_U$ violates at most $B$ of its constraints.
The path is called correct because, whenever it branches on an $(r-1)$-set $P$, it uses the branch labeled by the true assignment $x|_P$.
The forced values on $M_P^{x|_P}$ may still disagree with $x$; these disagreements are the errors we count below.

At a current non-base call with unassigned set $U$, let $\theta=\lceil B^{1/r}\rceil$.
For $P\in\binom U{r-1}$, define
\[
        e_P:=\#\{\,w\in M_P^{x|_P}:x_w\ne1-\beta_{P\cup\{w\}}(w)\,\}.
\]
This is exactly the number of variables that the branch $(P,x|_P)$ would set incorrectly.
If $w$ is counted in $e_P$, then $x$ agrees with the forbidden assignment on all of $P\cup\{w\}$, so the constraint on $P\cup\{w\}$ is violated by $x$.
Each violated $r$-set in the current restricted instance can be counted for at most its $r$ choices of $P$.
Hence, at every same-arity node on the path,
\[
        \sum_{P\in\binom U{r-1}} e_P\le rB .
\]

Call a branch $(P,x|_P)$ good if
\[
        |M_P^{x|_P}|\ge |U|/2^r
        \qquad\text{and}\qquad
        e_P\le\theta .
\]
Starting from the root, the correct path repeatedly follows any good branch, and stops when the current call is a base case or when no good branch exists.
Whenever the path follows a good branch, the branch is present in Algorithm~\ref{alg:det-close-list}; it fixes $P$ exactly as in $x$, introduces at most $\theta$ new disagreements on $M_P^{x|_P}$, and leaves at most $(1-2^{-r})|U|$ variables unassigned.
Therefore, after $t$ branching steps, the partial assignment fixed so far differs from $x$ in at most $t\theta$ positions, while the number of remaining variables is at most $(1-2^{-r})^t m$.
Thus $t=O(\log(m+2))$.

Let $U$ be the remaining variable set when the correct path stops.
The already fixed variables contribute at most $O(B^{1/r}\log(m+2))$ disagreements with $x$.
It remains to show that the algorithm adds a close completion on $U$.

\begin{itemize}
\item If the path stops at a base case, then the algorithm outputs all assignments on $U$, including $x|_U$.
This adds no further disagreement.

\item Suppose the path stops at a non-base call and $r=2$.
Let $z$ be the majority-type assignment added by Algorithm~\ref{alg:det-close-list} at this call, and write $e_u$ for $e_{\{u\}}$.
For any $u$ with $z_u\ne x_u$, the maximizing value in the definition of $z_u$ is $x_u$.
Since $|U|>4$ in a non-base call,
\[
        |M_u^{x_u}|\ge \frac{|U|-1}{2}\ge \frac{|U|}{4}.
\]
Because the path stopped, the branch $(u,x_u)$ is not good, and therefore $e_u>\theta$.
Using $\sum_{u\in U}e_u\le2B$,
\[
        \#\{u\in U:z_u\ne x_u\}
        \le \#\{u\in U:e_u>\theta\}
        \le \frac{2B}{\theta}
        =O(B^{1/2}).
\]
Thus extending the current partial assignment by $z$ gives distance $O(B^{1/2}\log(m+2))$ from $x$.

\item Suppose the path stops at a non-base call and $r\ge3$.
At this call, Algorithm~\ref{alg:det-close-list} performs its closing step: it chooses, for every $(r-1)$-set $P\subseteq U$, an assignment $\tau_P^*$ maximizing $|M_P^\tau|$, forms the complete one-forbidden $(r-1)$-CSP $\cK'$ that forbids $\tau_P^*$ on $P$, and calls $\mathrm{CloseList}(r-1,\cK',U,B')$ with $B'=\lceil rB/\theta\rceil$.

We claim that $x|_U$ violates at most $B'$ constraints of $\cK'$.
Indeed, a constraint $P$ of $\cK'$ is violated by $x|_U$ exactly when $x|_P=\tau_P^*$.
For such a $P$, maximality of $\tau_P^*$ gives
\[
        |M_P^{x|_P}|=|M_P^{\tau_P^*}|
        \ge \frac{|U|-r+1}{2^{r-1}}
        \ge \frac{|U|}{2^r},
\]
where the last inequality uses the non-base condition $|U|>2r$.
Since the correct path stopped at this call, the branch $(P,x|_P)$ is not good; the displayed size bound holds, so necessarily $e_P>\theta$.
Therefore
\[
\begin{aligned}
        \unsat_{\cK'}(x|_U)
        &=\#\{P\in\binom U{r-1}:x|_P=\tau_P^*\} \\
        &\le \#\{P\in\binom U{r-1}:e_P>\theta\}
        \le \frac{rB}{\theta}
        \le B'.
\end{aligned}
\]
By the induction hypothesis applied to $\cK'$ with budget $B'$, the lower-arity call contributes an assignment $y$ satisfying
\[
        \dist_H(y,x|_U)
        \le O\big((B')^{1/(r-1)}\log(m+2)\big)
        =O(B^{1/r}\log(m+2)),
\]
because $B'=O(B^{(r-1)/r})$ for fixed $r$.
Extending the current partial assignment by this $y$ gives the desired distance bound.
\end{itemize}

In all cases, the output list contains an assignment within $O(B^{1/r}\log(m+2))$ of $x$.
It remains to bound the list size and running time.
For fixed $r$, each same-arity non-base call has at most $2^{r-1}\binom{m}{r-1}=m^{O(1)}$ branching children, and every branching child reduces the current set size by a factor at most $1-2^{-r}$.
Thus there are $m^{O(\log m)}$ same-arity calls.
At each such call, the algorithm either adds one assignment when $r=2$, or makes one lower-arity call whose list size and running time are $m^{O(\log m)}$ by induction.
Therefore the total list size and running time are $m^{O(\log m)}$.
\end{proof}

\subsection{Randomized close-lists from direct products}
\label{sec:close-rand}
In this subsection, we present and analyze our randomized algorithm.
It calls the following set-indexed form of the direct-product decoder of Impagliazzo, Jaiswal, Kabanets, and Wigderson~\cite{IJKW10}.

\begin{theorem}[Set-indexed direct-product decoding]\label{thm:ijkw}
There is a universal constant $K_{\mathrm{IJKW}}$ such that the following holds.
Let $U$ be a set of size $m$, let $\ell\le m/2$, and let $Y$ be a function that maps each $W \in \binom{U}{\ell}$ to $Y(W)\in\{0,1\}^W$.
For an assignment $x\in\{0,1\}^U$, write $x^{(\ell)}(W)=x|_W$ and
\[
        \operatorname{Agr}(Y,x^{(\ell)}):=
        \Pr_{W\in\binom U\ell}[Y(W)=x|_W].
\]
If $\operatorname{Agr}(Y,x^{(\ell)})\ge\varepsilon$ and $\ell\ge K_{\mathrm{IJKW}}\log(1/\varepsilon)$, then a randomized algorithm with oracle access to $Y$ runs in time $\operatorname{poly}(m,1/\varepsilon)$ and outputs, with probability $\Omega(\varepsilon)$, an assignment $z\in\{0,1\}^U$ satisfying
\[
        \dist_H(z,x)\le
        K_{\mathrm{IJKW}}\frac{m\log(1/\varepsilon)}{\ell}.
\]
\end{theorem}

For an assignment $x\in\{0,1\}^V$, the map $x^{(\ell)}$ records the restriction $x|_W$ for every $\ell$-set $W\subseteq V$.
Theorem~\ref{thm:ijkw} says that if $Y$ gives a local assignment $Y(W)\in\{0,1\}^W$ for each $\ell$-set $W$, and if these local assignments agree with the restrictions of some global assignment $x$ on sufficiently many sets $W$, then one can recover a short list containing an assignment close to $x$.
In the proof below, we construct $Y$ from the induced subinstances on the sets $W$.
Thus Theorem~\ref{thm:ijkw} lets us turn local information from many small induced subinstances into candidate global assignments.

\begin{theorem}
\label{thm:random-close}
For every fixed $r\ge2$, given a complete Boolean $r$-CSP $\cI$ on a variable set $U$ of size $m$, in which every $r$-constraint forbids at least one local assignment, and a budget $B\ge1$, there is a randomized polynomial-time algorithm that outputs a polynomial-size set $\mathcal L\subseteq\{0,1\}^U$.
For every assignment $x$ with $\unsat_{\cI}(x)\le B$, with probability at least $2/3$ the set $\mathcal L$ contains an assignment $z$ satisfying
\[
        \dist_H(z,x)\le O(B^{1/r}\log(m+2)).
\]  
\end{theorem}

\begin{proof}
{\bf Parameters.}
\quad
Let 
\[
        \ell=\left\lfloor\frac{m}{2B^{1/r}}\right\rfloor,
        \qquad
        s_r(\ell) = \sum_{i=0}^{r-1} \binom{\ell}{i},
        \qquad
        \varepsilon_0=\frac{1}{4s_r(\ell)}, 
\]
so that $s_r(\ell)$ is the upper bound on the number of satisfying assignments on a complete $r$-CSP instance on $\ell$ variables by \Cref{lem:vc}.

One can assume $\ell = \omega(\log m)$, because otherwise, 
\[
m = O(B^{1/r} \log m),
\]
which implies that the trivial list $\mathcal{L} = \{ 0^U \}$ is a valid list. Since $\log(1/\eps_0) = O(r \log \ell)$, for sufficiently large $m$ (and $\ell$), we have $\ell \geq K_{\mathrm{IJKW}} \log(1/\eps_0)$ so that one can apply \cref{thm:ijkw} with $m, \ell$, and $\eps \leftarrow \eps_0$.

\paragraph{Algorithm.}
Our algorithm simply runs the decoder of \Cref{thm:ijkw} with input $\{ Y(W) \}_{W \in \binom{U}{\ell}}$, $O(1/\varepsilon_0)$ times, and outputs the union of all the decoded assignments. 
The input $Y$ is random and reused for every run of the decoder as follows. 
For a queried block $W\in\binom U\ell$, let $\cI[W]$ be the induced complete $r$-CSP on $W$, and let $\mathrm{Sat}(\cI[W])$ be its set of satisfying assignments.
This set can be enumerated by Lemma~\ref{lem:vc}.
If this set is nonempty, choose one of its assignments uniformly at random and set $Y(W)$ to that assignment.
If the set is empty, let $Y(W)$ be the all-zero assignment on $W$.

The algorithm implements the same distribution lazily, sampling $Y(W)$ the first time the decoder queries $W$ and reusing the sample thereafter.
For the analysis, sample all values $Y(W)$ independently in advance.

\paragraph{Analysis.}
Fix $x$ with $\unsat_{\cI}(x)\le B$.
For a uniformly random block $W$, each violated $r$-set of $x$ is contained in $W$ with probability $\binom{\ell}{r}/\binom{m}{r}\le (\ell/m)^r$.
Hence
\[
        \ex[\unsat_{\cI[W]}(x|_W)]
        \le B\left(\frac{\ell}{m}\right)^r\le 2^{-r}\le\frac14.
\]
Thus $x|_W$ satisfies $\cI[W]$ with probability at least $3/4$.
Conditional on this event, the oracle value equals $x|_W$ with probability at least $1/s_r(\ell)$.
Therefore
\[
        \ex_Y[\operatorname{Agr}(Y,x^{(\ell)})]
        \ge \frac{3}{4s_r(\ell)}=3\varepsilon_0.
\]
The agreement indicators are independent over the blocks, though not necessarily identically distributed.
Since their average has expectation at least $3\varepsilon_0$, the multiplicative Chernoff bound gives, for sufficiently large $m$, 
\[
        \Pr_Y[\operatorname{Agr}(Y,x^{(\ell)})<\varepsilon_0]
        \le \exp\left(-\frac{2}{3}\varepsilon_0\binom m\ell\right)
        \le \frac{1}{10}.
\]
Condition on the event $\operatorname{Agr}(Y,x^{(\ell)})\ge\varepsilon_0$.
Theorem~\ref{thm:ijkw} applies with $\varepsilon=\varepsilon_0$.
One run returns an assignment at distance
\[
        K_{\mathrm{IJKW}}\frac{m\log(1/\varepsilon_0)}{\ell}
        =O(B^{1/r}\log(m+2))
\]
from $x$ with probability $\Omega(\varepsilon_0)$.
Repeating the decoder $O(1/\varepsilon_0)$ times raises the conditional success probability to at least $9/10$.
The overall success probability is at least $(9/10)^2>2/3$.
Since $1/\varepsilon_0=O(\ell^{r-1})$ and $r$ is fixed, the number of decoder calls and oracle queries is polynomial in $m$.
\end{proof}

\subsection{Exposing low-arity residuals}
\label{sec:pivot}

In this subsection, we prove the main theorem of this section, \Cref{thm:pivot}. 
Given a Min-CSP($A$) instance $\cI$ on the variable set $V$ with arity $k$, let $r := \max(2, w(A))$ and $q := k - r$. 

The following lemma shows that after fixing any set of $k-r$ variables, each residual $r$-ary constraint still forbids at least one local assignment.

\begin{lemma}[Residual nontriviality from width]\label{lem:residual}
For any $P \subseteq V$ with $|P| = q$ and a partial assignment $x \in \{ 0, 1 \}^P$, consider the complete $r$-CSP instance $\cJ$ on the variable set $V \setminus P$ such that, given an assignment $y \in \{ 0, 1\}^{V \setminus P}$ and a set $W \subseteq V \setminus P$, the constraint for $W$ is satisfied if $(x|_P \cup y|_W)$ satisfies the constraint for $P \cup W$ of $\cI$. Then, every $r$-ary constraint forbids at least one local assignment.
\end{lemma}

\begin{proof}
Consider an arbitrary $W \subseteq V \setminus P$ and suppose the given partial assignment $x$ to $P$ contributes $t$ true literals after applying negations to the constraint for $P \cup W$ in $\cI$.
When the remaining $r$ literals vary, their contribution can be any value in $\{0,1,\ldots,r\}$.
If the residual constraint for $W$ in $\cJ$ accepted every local assignment, then
\[
        \{t,t+1,\ldots,t+r\}\subseteq A,
\]
which is a consecutive accepted interval of length $r+1$. Since $r \geq w(A)$, it contradicts the definition of $w(A)$. 
\end{proof}

Given the above lemma, our main \Cref{thm:pivot} follows by applying the algorithms from the previous subsections to these residual instances.

\begin{theorem}[Close-lists after exposing a pivot]\label{thm:pivot}
Let $r$ be as in Lemma~\ref{lem:residual}, and put $q=k-r$.
There is a constant $K_{\mathrm{list}}$ such that, for every budget $T\ge1$, if
\[
        H_T:=\min\left\{n,
        \left\lceil K_{\mathrm{list}}
        \left(\max\left\{1,\frac{T}{n^q}\right\}\right)^{1/r}\log n
        \right\rceil\right\},
\]
then the following deterministic and randomized close-list algorithms exist.
\begin{enumerate}
\renewcommand{\labelenumi}{(\alph{enumi})}
\item A deterministic algorithm constructs a list $Z_T\subseteq\{0,1\}^V$ of size $n^{O(\log n)}$ in time $n^{O(\log n)}$ such that, for every assignment $x$ with $\unsat_{\cI}(x)\le T$, the list contains $z\in Z_T$ satisfying $\dist_H(z,x)\le H_T$.
\item A randomized polynomial-time algorithm outputs a polynomial-size set $Z_T\subseteq\{0,1\}^V$ such that, for every fixed assignment $x$ with $\unsat_{\cI}(x)\le T$, the set contains an assignment $z$ satisfying $\dist_H(z,x)\le H_T$ with probability at least $2/3$.
\end{enumerate}
\end{theorem}

\begin{proof}
For a $q$-set $P\subseteq V$, let $d_P(x)$ denote the number of violated constraints whose scope contains $P$.
Double-counting violated constraints gives
\[
        \sum_{P\in\binom Vq} d_P(x)=\binom kq \unsat_{\cI}(x)\le \binom kq T.
\]
Since $n\ge k$ and $q$ is fixed, some $q$-set $P$ satisfies
\[
        d_P(x)\le O(T/n^q).
\]

Both algorithms try all choices of $P$ and all assignments $b\in\{0,1\}^P$.
For a fixed choice, each $r$-set $R\subseteq V\setminus P$ inherits the residual constraint on $R$ obtained from the original constraint on $P\cup R$ after setting $P=b$.
Lemma~\ref{lem:residual} says that every residual constraint forbids at least one local assignment.

The deterministic algorithm chooses one forbidden local assignment for every residual $r$-set and applies Theorem~\ref{thm:oneforbidden-list} with budget
\[
        B=\left\lceil C\max\left\{1,T/n^q\right\}\right\rceil,
\]
where $C=C(k)$ is a large enough constant.
For the pivot $P$ from the averaging argument and $b=x|_P$, the corresponding instance has at most $d_P(x)$ violations under $x|_{V\setminus P}$.
After extending by $b$ on $P$, this list contributes an assignment in $Z_T$ within $O(B^{1/r}\log n)$ of $x$.

The randomized algorithm keeps the full residual $r$-CSP and applies Theorem~\ref{thm:random-close} with the same budget $B$.
For the same pivot $P$ and assignment $b=x|_P$, the residual instance has at most $d_P(x)$ violations under $x|_{V\setminus P}$.
With probability at least $2/3$, the resulting set $Z_T$ contains some $z$ within $O(B^{1/r}\log n)$ of $x$.

Taking $K_{\mathrm{list}}$ large enough as a function of $k$, the corresponding distances are at most $H_T$.
The deterministic list has size $n^{O(\log n)}$, while the randomized output is polynomial-size, because $q$ and $r$ are fixed.
\end{proof}

\subsection{Exact enumeration at very small optimum}
\label{sec:smallopt}

In this subsection, we prove the following theorem for the very small optimum case.
The idea is that one residual instance has optimum zero, and Lemma~\ref{lem:vc} can enumerate all satisfying assignments of that residual instance. 

\begin{theorem}[Exact enumeration for very small optimum]\label{thm:smallopt}
With $r$ as in Lemma~\ref{lem:residual} and $q=k-r$, there is a threshold $\tau \in (0, 1)$ and, for fixed $k$, a deterministic polynomial-time procedure that returns an optimum assignment whenever
\[
        \opt(\cI)<\tau n^q.
\]
\end{theorem}

\begin{proof}
Let $x^*$ be an optimum assignment.
If $q=0$, let $\tau = 1/2$ so that the condition 
$\opt(\cI)<\tau n^q$ implies
$\opt(\cI)=0$. Lemma~\ref{lem:vc} applied directly to the complete $k$-CSP enumerates all satisfying assignments, so the optimum is found.

For $q>0$, let
\[
        \tau=\frac{1}{2\binom kq q^q}.
\]
Using the notation from the proof of Theorem~\ref{thm:pivot},
\[
        \frac{1}{\binom nq}\sum_{P\in\binom Vq}d_P(x^*)
        =\frac{\binom kq\opt(\cI)}{\binom nq}
        < 1,
\]
using $\binom nq\ge (n/q)^q$ and $\opt(\cI)<\tau n^q$.
Thus some $P$ has $d_P(x^*)=0$.

The procedure enumerates all $q$-sets $P$ and all assignments $b$ on $P$.
For each choice it forms the complete residual $r$-CSP on $V\setminus P$ coming from the constraints on $P\cup R$ after setting $P=b$.
Lemma~\ref{lem:residual} says that every residual constraint forbids at least one local assignment, so Lemma~\ref{lem:vc} enumerates all satisfying assignments of this residual instance.
Each is extended by $b$ and evaluated in the original objective.

For the zero-violation pivot $P$ and the matching assignment $b=x^*|_P$, the residual enumeration contains $x^*|_{V\setminus P}$.
Hence the best evaluated extension is optimum for the original instance.
\end{proof}

\section[Approximation for width at least three]{Approximation for $w(A)\ge3$}
\label{sec:w-ge-3}

Throughout this section $3\le w(A)\le k$.
We use the randomized part of Theorem~\ref{thm:pivot} to obtain an assignment $z$ close to an optimum assignment.
From this reference assignment, the rounding step encodes constraints with one or two flipped variables as a weighted unary/binary Min-2-SAT instance.
Constraints with at least three flipped variables are handled by a direct count.

\subsection{Rounding from a nearby assignment}

In this subsection, we obtain an approximation guarantee starting from a given assignment that is close to the optimal assignment. The key idea is a reduction to Weighted Min-2-SAT.

\begin{theorem}[Weighted Min-2-SAT~\cite{ACMM05}]\label{thm:min2sat}
A weighted Boolean instance consisting of unary and binary constraints, each forbidding one local assignment and having nonnegative integer weight, admits a randomized polynomial-time $O(\sqrt{\log n})$ approximation, where $n$ is the number of variables.
The success probability can be made at least $5/6$ by constant repetition and keeping the lowest-cost output.
\end{theorem}

Fix an assignment $z\in\{0,1\}^V$.
A correction is a vector $s\in\{0,1\}^V$, where $s_v=1$ means that $v$ is flipped from $z$.
For such a correction, $z\oplus s$ denotes the assignment with $(z\oplus s)_v=z_v\oplus s_v$ for every $v\in V$.
If $\eta\in\{0,1\}^S$ is a correction only on a set $S$, then $z\oplus\eta$ means that variables in $S$ are flipped according to $\eta$ and variables outside $S$ are not flipped.
For $W\subseteq V$, let $0_W$ denote the all-zero assignment on $W$.
For a $k$-set $S$, define
\[
        \mathcal B_S(z):=\{\eta\in\{0,1\}^S:
        \text{the constraint on }S\text{ is violated by }z\oplus\eta\}.
\]
Thus $z\oplus s$ violates $S$ exactly when $s|_S\in \mathcal B_S(z)$.

\begin{lemma}[Rounding from a nearby assignment]\label{lem:auxiliary}
Let $x^*$ be an optimum assignment.
Suppose that, for some nonnegative integer $H$, we are given an assignment $z$ satisfying $\dist_H(z,x^*)\le H$ and
\[
        \opt(\cI)\le n^{k-3}H^3.
\]
Then there is a randomized polynomial-time algorithm that, with probability at least $5/6$, returns an assignment $\widehat x$ satisfying
\[
        \unsat_{\cI}(\widehat x)\le O(n^{k-3}H^3\log^{3/2} n).
\]
\end{lemma}

\begin{proof}
If $H=0$, then $z=x^*$ and the hypothesis gives $\opt(\cI)=0$, so returning $z$ suffices.
Assume $H\ge1$.
Let
\[
        R^*=\{v:z_v\ne x^*_v\},
        \qquad |R^*|\le H,
\]
and let $s^*=1_{R^*}$ be the optimum correction. We construct a Weighted Min-2-SAT instance on the variable set $V$ as follows. 

\paragraph{Reduction.}
Set
\[
        \Lambda=H^2n^{k-3},
        \qquad
        \Delta=Hn^{k-3}.
\]
The weight $\Lambda$ penalizes the number of flips.
The parameter $\Delta=Hn^{k-3}$ is chosen so that, for any fixed pair $P$, at most $|R^*|\binom{n}{k-3}\le Hn^{k-3}=\Delta$ sets $W\in\binom{V\setminus P}{k-2}$ intersect $R^*$. 
There are the following three types of constraints.

\begin{itemize}
\item For every variable $v$, add a unary constraint forbidding $s_v=1$ with weight $\Lambda$.
\item For every pair $P=\{u,v\}$ and local assignment $\tau\in\{0,1\}^P$, let
\[
        \mathcal N_P^\tau=
        \left\{W\in\binom{V\setminus P}{k-2}:
        \tau\cup 0_W\in \mathcal B_{P\cup W}(z)\right\},
        \qquad
        b_P^\tau=\max\{0,|\mathcal N_P^\tau|-\Delta\}.
\]
If $b_P^\tau>0$, add a binary constraint forbidding $s|_P=\tau$ with weight $b_P^\tau$.
\item 
For $v\in V$ and $a\in\{0,1\}$, let $\mathcal H_v^a$ be the $(k-1)$-uniform hypergraph on $V\setminus\{v\}$ whose edges are the sets $W$ with
\[
        (v\mapsto a)\cup 0_W\in \mathcal B_{W\cup\{v\}}(z).
\]
A vertex of $\mathcal H_v^a$ is called heavy if its degree is larger than $4\Delta$.
Put
\[
        u_v^a=\max\{0,\#\{W\in E(\mathcal H_v^a):W
        \text{ contains no heavy vertex}\}-4H\Delta\}.
\]
If $u_v^a>0$, add a unary constraint forbidding $s_v=a$ with weight $u_v^a$.
\end{itemize}
Then, for any correction $s$, its cost in the Weighted Min-2-SAT instance is
\[
        \Psi(s):=\Lambda|\{v:s_v=1\}|+
        \sum_{P\in\binom V2} b_P^{s|_P}+
        \sum_{v\in V} u_v^{s_v}.
\]
We call the second sum the \emph{binary part} and the third sum the \emph{unary part}.

\paragraph{Completeness. }
First, we bound the cost of the true correction $\Psi(s^*)$.
The cardinality term is at most $\Lambda H=n^{k-3}H^3$.

For the binary part, fix a pair $P$. At most $\Delta$ members of
$\mathcal N_P^{s^*|_P}$ meet $R^*$, and hence
\[
        b_P^{s^*|_P}
        =\max\{0,|\mathcal N_P^{s^*|_P}|-\Delta\}
        \le
        \bigl|\{W\in\mathcal N_P^{s^*|_P}:W\cap R^*=\emptyset\}\bigr|.
\]
For every $W\in\mathcal N_P^{s^*|_P}$ with $W\cap R^*=\emptyset$, the
correction on $P\cup W$ agrees with $s^*$, so the constraint on $P\cup W$
is violated by $x^*$. Each violated $k$-set is counted for at most
$\binom{k}{2}$ choices of $P$. Therefore
\[
        \sum_{P\in\binom V2} b_P^{s^*|_P}
        \le \binom{k}{2}\opt(\cI) .
\]

For the unary part, fix $v$.
Every set $W$ counted below contains no heavy vertex.
Hence, if it also intersects $R^*$, then it contains some non-heavy vertex of $R^*\setminus\{v\}$.
Since every non-heavy vertex has degree at most $4\Delta$ in $\mathcal H_v^{s_v^*}$,
\[
\begin{aligned}
        &\bigl|\{W\in E(\mathcal H_v^{s_v^*}):
        W \text{ contains no heavy vertex and } W\cap R^*\ne\emptyset\}\bigr| \\
        &\qquad\le
        \sum_{\substack{y\in R^*\setminus\{v\}\\
        y\text{ is non-heavy in }\mathcal H_v^{s_v^*}}}
        \deg_{\mathcal H_v^{s_v^*}}(y)
        \le 4H\Delta .
\end{aligned}
\]
Thus
\[
        u_v^{s_v^*}
        \le
        \bigl|\{W\in E(\mathcal H_v^{s_v^*}):
        W \text{ contains no heavy vertex and } W\cap R^*=\emptyset\}\bigr|.
\]
For every such $W$, the correction on $W\cup\{v\}$ agrees with $s^*$, so
the constraint on $W\cup\{v\}$ is violated by $x^*$. Each violated $k$-set
is counted for at most $k$ choices of $v$. Hence
\[
        \sum_{v\in V} u_v^{s_v^*}\le k\opt(\cI) .
\]
Consequently, using $\opt(\cI)\le n^{k-3}H^3$,
\[
        \Psi(s^*)
        \le \Lambda H+\left(\binom{k}{2}+k\right)\opt(\cI)
        =O(n^{k-3}H^3).
\]
\paragraph{Soundness.}
With probability at least $5/6$, Theorem~\ref{thm:min2sat} returns a correction $\widehat s$ whose cost is within an $O(\sqrt{\log n})$ factor of optimum for the weighted Min-2-SAT instance.
Using the completeness bound for $s^*$, on this event we have
\[
        \Psi(\widehat s)\le O(n^{k-3}H^3\sqrt{\log n}),
        \qquad
        |\{v:\widehat s_v=1\}|\le \frac{\Psi(\widehat s)}{\Lambda}
        \le O(H\sqrt{\log n}).
\]
Put $\widehat R=\{v:\widehat s_v=1\}$ and $F=R^*\cup\widehat R$, so $|F|\le O(H\sqrt{\log n})$.
Outside $F$, the assignments $z\oplus\widehat s$ and $x^*$ agree.
Therefore constraints disjoint from $F$ contribute at most $\opt(\cI)$ violations, and it remains to count violated constraints meeting $F$.

If a constraint violated by $z\oplus\widehat s$ has scope $S$ with $|S\cap F|\ge3$, then there are at most $O(n^{k-3}|F|^3)$ choices for $S$.
If a constraint violated by $z\oplus\widehat s$ has scope $S$ with $|S\cap F|=2$, write $S=P\cup W$ with $P\in\binom F2$ and $W\cap F=\emptyset$.
For a fixed pair $P$, every such $W$ lies in $\mathcal N_P^{\widehat s|_P}$, because all variables of $W$ have correction value $0$ under $\widehat s$.
By the definition of $b_P^{\widehat s|_P}$, the number of such $W$ is at most $b_P^{\widehat s|_P}+\Delta$.
Hence all violated constraints whose scopes contain exactly two vertices of $F$ contribute at most
\[
        \sum_{P\in\binom V2} b_P^{\widehat s|_P}+\Delta |F|^2,
\]
where the sum is the binary part of $\Psi(\widehat s)$.

Consider constraints violated by $z\oplus\widehat s$ whose scopes meet $F$ in exactly one vertex, say $S=\{v\}\cup W$ with $v\in F$ and $W\cap F=\emptyset$.
For a fixed $v$, the light choices of $W$ are exactly the edges of $\mathcal H_v^{\widehat s_v}$ that contain no heavy vertex.
By the definition of $u_v^{\widehat s_v}$, their number is at most $u_v^{\widehat s_v}+4H\Delta$.
After summing over $v\in F$ and upper-bounding by the full unary part, these light choices contribute at most
\[
        \sum_{v\in V} u_v^{\widehat s_v}+4H\Delta |F|,
\]
where the sum is the unary part of $\Psi(\widehat s)$.

It remains to count violated constraints whose scopes meet $F$ in exactly one vertex and whose corresponding set $W$ contains a heavy vertex.
Orient each such constraint, with scope $S=\{v\}\cup W$, to one heavy vertex $y\in W$.
Since $W\cap F=\emptyset$, we have $y\notin F$ and $\widehat s_y=0$.
For a fixed ordered pair $(v,y)$, let $\tau_{v,y}$ be the local assignment on $\{v,y\}$ with $\tau_{v,y}(v)=\widehat s_v$ and $\tau_{v,y}(y)=0$.
Deleting $v$ and $y$ from the scope of each oriented constraint gives a set in $\mathcal N_{\{v,y\}}^{\tau_{v,y}}$.
This map is injective for fixed $(v,y)$, because the original scope is recovered by adding $v$ and $y$ back.
Moreover, because $y$ is heavy in $\mathcal H_v^{\widehat s_v}$,
\[
        |\mathcal N_{\{v,y\}}^{\tau_{v,y}}|>4\Delta.
\]
Therefore
\[
        b_{\{v,y\}}^{\tau_{v,y}}
        =|\mathcal N_{\{v,y\}}^{\tau_{v,y}}|-\Delta
        \ge \frac34|\mathcal N_{\{v,y\}}^{\tau_{v,y}}|,
\]
and hence the number of heavy constraints charged to this ordered pair is at most
\[
        |\mathcal N_{\{v,y\}}^{\tau_{v,y}}|
        \le 2b_{\{v,y\}}^{\tau_{v,y}}.
\]
Each charged unordered pair has exactly one endpoint in $F$, namely $v$; the other endpoint $y$ is outside $F$.
Thus the same binary weight cannot be charged again with the opposite orientation.
All heavy constraints in the one-intersection case therefore contribute at most twice the binary part of $\Psi(\widehat s)$.

Combining the three cases,
\[
        \unsat_{\cI}(z\oplus\widehat s)
        \le \opt(\cI)+O\left(\Psi(\widehat s)+H\Delta |F|+
        \Delta |F|^2+n^{k-3}|F|^3\right).
\]
Using $\Delta=O(Hn^{k-3})$, $|F|\le O(H\sqrt{\log n})$, $\Psi(\widehat s)\le O(n^{k-3}H^3\sqrt{\log n})$, and $\opt(\cI)\le n^{k-3}H^3$, the right-hand side is at most $O(n^{k-3}H^3\log^{3/2} n)$.
Returning $\widehat x=z\oplus\widehat s$ proves the lemma on the success event of Theorem~\ref{thm:min2sat}.
\end{proof}

\subsection[The width-at-least-three algorithm]{The algorithm for $w(A)\ge3$}

In this subsection, we combine the close-list theorem with the rounding lemma to obtain the approximation guarantee for complete Min-CSP($A$) when $w(A) \ge 3$.

\begin{theorem}\label{thm:wge3}
If $3\le w(A)\le k$, there is a randomized polynomial-time algorithm that succeeds with probability at least $2/3$ and returns an assignment $\widehat x$ with
\[
        \unsat_{\cI}(\widehat x)\le O(n^{w(A)-3}\log^{9/2} n)\,\opt(\cI).
\]
\end{theorem}

\begin{proof}
Put $r:=w(A)$ and $q:=k-r$.

\paragraph{Algorithm.}
Maintain a set $\mathcal C$ of candidate assignments, initially containing an arbitrary assignment.
First run the exact small-optimum procedure of Theorem~\ref{thm:smallopt}; if it outputs an assignment, add it to $\mathcal C$.

Then consider all dyadic budgets
\[
        T=1,2,4,\ldots,2^{\lceil\log_2\binom nk\rceil}.
\]
For each such $T$, run the randomized close-list algorithm of Theorem~\ref{thm:pivot}(b) twice independently and let $Z_T$ be the union of the two output lists.
For every $z\in Z_T$, run Lemma~\ref{lem:auxiliary} with reference assignment $z$ and radius $H_T$, where $H_T$ is the radius from Theorem~\ref{thm:pivot} for this budget $T$.
Add every assignment returned by Lemma~\ref{lem:auxiliary} to $\mathcal C$.
Finally, return an assignment in $\mathcal C$ minimizing $\unsat_{\cI}$.

\paragraph{Analysis.}
Let $x^*$ be an optimum assignment.
If $\opt(\cI)<\tau n^{k-w(A)}$, where $\tau$ is the threshold from Theorem~\ref{thm:smallopt}, that theorem already returns an optimum assignment.
Otherwise choose a dyadic $T$ with $\opt(\cI)\le T<2\opt(\cI)$.
With probability at least $8/9$, the two independent close-list calls for this budget produce a list $Z_T$ containing an assignment $z$ with $\dist_H(z,x^*)\le H_T$, where $H_T$ was defined as 
\[
        H_T = \min\left\{n,
        \left\lceil K_{\mathrm{list}}
        \left(\max\left\{1,\frac{T}{n^q}\right\}\right)^{1/r}\log n
        \right\rceil\right\}.
\]
Condition on this event.

To apply Lemma~\ref{lem:auxiliary} to this assignment $z$, first check its hypothesis.
If $H_T=n$, then $n^{k-3}H_T^3=n^k\ge T\ge\opt(\cI)$.
Here $k\ge3$ and the chosen budget satisfies $T<2\opt(\cI)\le2\binom nk\le n^k$.
If $H_T < n$, then
\[
        H_T\ge K_{\mathrm{list}}
        \left(\max\left\{1,\frac{T}{n^{k-w(A)}}\right\}\right)^{1/w(A)}\log n.
\]
When $T\ge n^{k-w(A)}$, this gives
\[
        \frac{n^{k-3}H_T^3}{T}
        \ge K_{\mathrm{list}}^3\log^3 n
        \left(\frac{n^k}{T}\right)^{(w(A)-3)/w(A)}\ge 1.
\]
Here $K_{\mathrm{list}}\ge2$, $n\ge k\ge3$, and $T<2\opt(\cI)\le2\binom nk\le n^k$.
When $T<n^{k-w(A)}$, the same bound gives $H_T\ge K_{\mathrm{list}}\log n\ge1$.
Since $k-w(A)\le k-3$, we again have $n^{k-3}H_T^3\ge T$.
Thus $\opt(\cI)\le T\le n^{k-3}H_T^3$, so Lemma~\ref{lem:auxiliary} applies.

The same definition of $H_T$ gives
\[
        H_T\le O\left(
        \left(\max\left\{1,\frac{T}{n^{k-w(A)}}\right\}\right)^{1/w(A)}\log n
        \right).
\]
Since $T\ge \tau n^{k-w(A)}$, we have $\max\left\{1,\frac{T}{n^{k-w(A)}}\right\} = O(T/n^{k-w(A)})$.
Therefore
\[
\begin{aligned}
        n^{k-3}H_T^3
        &\le O\left(n^{k-3}
        \left(\frac{T}{n^{k-w(A)}}\right)^{3/w(A)}\log^3 n\right) \\
        &= O\left(n^{w(A)-3}T
          \left(\frac{n^{k-w(A)}}{T}\right)^{(w(A)-3)/w(A)}\log^3 n\right) \\
        &\le O(n^{w(A)-3}T\log^3 n).
\end{aligned}
\]
Since $T<2\opt(\cI)$, Lemma~\ref{lem:auxiliary} returns, with probability at least $5/6$ conditional on the close-list event, an assignment of cost at most $O(n^{w(A)-3}\log^{9/2} n)\,\opt(\cI)$.
Thus the candidate set contains such an assignment with probability at least $(8/9)(5/6)>2/3$, and the algorithm returns one whenever this happens.
All close lists have polynomial size, all rounding calls run in polynomial time, and the number of budgets is $O(\log n)$.
Hence the algorithm runs in randomized polynomial time.
\end{proof}

\section[QPTAS for width at most two]{QPTAS for $w(A)\le 2$}
\label{sec:w-le-2}

Throughout this section $1\le w(A)\le2$.
We apply the deterministic part of Theorem~\ref{thm:pivot} with $r=2$ and $q=k-2$.
For the budget within a constant factor of the optimum, the list contains an assignment $z$ such that
\[
        \dist_H(z, x^*) = O\left(\left(\frac{\opt(\cI)}{n^{k-2}}\right)^{1/2}\log n\right)
\]
where $x^*$ is an optimum assignment.
Around such an assignment, the objective has a degree-$k$ expansion in the flipped variables.
The terms of degree at least three are negligible in the low-optimum range; the quadratic part is handled by the search lemma below.

\subsection{Expansion around a reference assignment}

We first separate the objective into its quadratic part and a controlled higher-degree remainder.
For an assignment $z$ and a set $R\subseteq V$, write $z\oplus R$ for the assignment obtained from $z$ by flipping exactly the variables in $R$.
\begin{lemma}[Bounded coefficients]\label{lem:polyexpansion}
Fix an assignment $z\in\{0,1\}^V$.
There are unique coefficients $a_D$, indexed by nonempty sets $D\subseteq V$ of size at most $k$, such that for every set $R\subseteq V$ of variables to flip,
\[
        \unsat_{\cI}(z\oplus R)=\unsat_{\cI}(z)+
        \sum_{d=1}^k\sum_{D\in\binom Vd} a_D\prod_{i\in D}1[i\in R].
\]
They satisfy
\[
        |a_D|=O(n^{k-|D|}).
\]
Writing
\[
        Q_z(R)=\sum_{i\in R}a_{\{i\}}+
        \sum_{\{i,j\}\subseteq R}a_{\{i,j\}},
        \qquad
        E_z(R)=\sum_{d=3}^k\sum_{\substack{D\subseteq R\\ |D|=d}}a_D,
\]
we have
\[
        \unsat_{\cI}(z\oplus R)=\unsat_{\cI}(z)+Q_z(R)+E_z(R),
\]
and, whenever $|R|\le H$,
\[
        |E_z(R)|\le O(n^{k-3}H^3).
\]
\end{lemma}

\begin{proof}
For a single constraint $S$, the violation indicator as a function of the flip bits on $S$ has a unique multilinear expansion of degree at most $k$, with coefficients bounded by a constant depending only on $k$.
The global coefficient $a_D$ receives contributions only from constraints whose scope contains $D$.
There are $\binom{n-|D|}{k-|D|}=O(n^{k-|D|})$ such scopes, proving the coefficient bound.

Summing the single-constraint expansions gives the identity.
For the remainder, since $R\subseteq V$,
\[
        |E_z(R)|
        \le O\left(\sum_{d=3}^k n^{k-d}|R|^d\right)
        \le O(n^{k-3}|R|^3)
        \le O(n^{k-3}H^3),
\]
because $k$ is fixed and $|R|\le n$.
\end{proof}

\subsection{Quadratic minimization with bounded pair weights}

The QPTAS uses a fixed-cardinality quadratic minimization primitive for possibly negative pair weights and arbitrary linear terms.
We first prove a truncated-simplex enumeration lemma using the $\ell_\infty$ approximate Carath\'eodory bound stated by Barman~\cite{Barman18}.
Lemma~\ref{lem:quad} then rounds the fractional solution to an integral set.

\begin{lemma}[Quadratic search on a truncated simplex]\label{lem:quad-simplex}
Let
\[
        \Delta_t=\left\{p\in\mathbb R_{\ge0}^n:\sum_i p_i=1,\ p_i\le \frac1t
        \text{ for all }i\right\},
\]
where $2\le t\le n$.
Let $C$ be an $n\times n$ polynomial-bit rational matrix with $|C_{ij}|\le1$, and let $d\in\mathbb Q^n$ have polynomial bit complexity.
For every $\eta\in(0,1)$ there is a deterministic algorithm running in time $n^{O(\log n/\eta^2)}$ that returns $p\in\Delta_t$ with
\[
        d\cdot p+p^\transpose C p
        \ge \max_{q\in\Delta_t}\left(d\cdot q+q^\transpose Cq\right)-\eta .
\]
\end{lemma}

\begin{proof}
Let $p^*$ maximize the displayed objective over $\Delta_t$.
For $i=1,\ldots,n$, let $C_i\in\mathbb R^n$ be the $i$th column of $C$.
Then $\|C_i\|_\infty\le1$ for every $i$, and
\[
        Cp^*=\sum_{i=1}^n p^*_i C_i
\]
belongs to $\operatorname{conv}\{C_1,\ldots,C_n\}$.
We use the $\ell_\infty$ approximate Carath\'eodory bound stated by Barman~\cite{Barman18}: for every point in the convex hull of $n$ vectors of $\ell_\infty$-norm at most one, there is an average of $O(\log n/\delta^2)$ of those vectors, with repetitions allowed, within $\ell_\infty$-distance $\delta$.
Choose a sufficiently large universal constant $C_B$ so that, with $\delta=\eta/4$, the theorem guarantees an average of at most
\[
        s:=\left\lceil C_B\frac{\log(2n)}{\eta^2}\right\rceil
\]
columns within $\ell_\infty$-distance $\eta/4$ of $Cp^*$.
Thus there are indices $i_1,\ldots,i_r$, with $1\le r\le s$, such that
\[
        u:=\frac1r\sum_{h=1}^r C_{i_h}
        \qquad\text{satisfies}\qquad
        \|u-Cp^*\|_\infty\le \eta/4.
\]

\paragraph{Algorithm.}
The algorithm enumerates all index sequences of length at most $s$.
For each resulting vector $u$, it solves the linear program
\[
\begin{array}{ll}
\text{maximize}   & d\cdot p+p\cdot u \\
\text{subject to} & p\in\Delta_t, \\
                  & -\eta/4\le (Cp-u)_j\le\eta/4
                    \quad\text{for every }j=1,\ldots,n.
\end{array}
\]
If the linear program is infeasible, the algorithm discards this $u$.
Among all feasible outputs, it returns the point with largest true objective $d\cdot p+p^\transpose Cp$.

\paragraph{Analysis.}
Consider the vector $u$ guaranteed above.
Then $p^*$ is feasible for its linear program.
If $p$ is an optimum solution of that program, then
\[
        d\cdot p+p\cdot u\ge d\cdot p^*+p^*\cdot u.
\]
Moreover, since $\|p\|_1=\|p^*\|_1=1$, since $p$ satisfies the linear program constraints, and since $\|u-Cp^*\|_\infty\le\eta/4$,
\[
        |p^\transpose(Cp-u)|\le\eta/4,
        \qquad
        |(p^*)^\transpose(Cp^*-u)|\le\eta/4.
\]
Therefore
\[
\begin{aligned}
        d\cdot p+p^\transpose Cp
        &\ge d\cdot p+p\cdot u-\eta/4 \\
        &\ge d\cdot p^*+p^*\cdot u-\eta/4 \\
        &\ge d\cdot p^*+(p^*)^\transpose Cp^*-\eta/2 \\
        &\ge d\cdot p^*+(p^*)^\transpose Cp^*-\eta .
\end{aligned}
\]
Since the algorithm returns the feasible output with largest true objective, the returned point satisfies the same bound.
Finally, $s=O(\log n/\eta^2)$, so enumerating all sequences of length at most $s$ and solving the corresponding linear programs takes time $n^{O(\log n/\eta^2)}$.
\end{proof}

The next lemma uses the truncated-simplex search to minimize a quadratic set function over sets of size at most $H$.

\begin{lemma}[Bounded quadratic rounding]\label{lem:quad}
Let
\[
        F(R)=\sum_{i\in R}a_i+\sum_{\{i,j\}\subseteq R}b_{ij},
        \qquad |b_{ij}|\le B,
\]
where the linear coefficients are arbitrary and $B>0$.
For polynomial-bit rational input coefficients, given $H\in\{0,1,\ldots,n\}$ and $\eta\in(0,1)$, there is a deterministic algorithm running in time
\[
        n^{O(\log n/\eta^2)}
\]
that returns $R\subseteq V$, $|R|\le H$, satisfying
\[
        F(R)\le \min_{|R'|\le H}F(R')+\eta BH^2.
\]
\end{lemma}

\begin{proof}
It is enough to handle each cardinality $t\le H$ and return the best set found.
The cases $t=0$ and $t=1$ are solved exactly, so fix $t\ge2$.

Put $C_{ii}=1$, $C_{ij}=-b_{ij}/(2B)$ for $i\ne j$, and $d_i=-a_i/(Bt)$.
Then $|C_{ij}|\le1$ for all $i,j$, and in fact $|C_{ij}|\le1/2$ for $i\ne j$.
Apply Lemma~\ref{lem:quad-simplex} with accuracy $\eta/4$ to maximize
\[
        g(p)=d\cdot p+p^\transpose C p
\]
over $\Delta_t$.
For a $t$-set $R$, let $p_R=1_R/t$.
A direct calculation gives
\[
        g(p_R)= -\frac{F(R)}{Bt^2}+\frac1t .
\]
Thus, up to the constant $1/t$, maximizing $g$ over integral points is the same as minimizing $F$ over $t$-sets.

Round the fractional point returned by Lemma~\ref{lem:quad-simplex} without decreasing $g$.
If two coordinates $i,j$ are fractional, vary them as $p+\delta(e_i-e_j)$ over the maximal interval that keeps the point in $\Delta_t$.
The function of $\delta$ is a convex quadratic, because its quadratic coefficient is
\[
        C_{ii}+C_{jj}-2C_{ij}=2-2C_{ij}\ge1 .
\]
Therefore one endpoint has value at least the current value.
Moving to such an endpoint makes one additional coordinate integral.
Repeating this operation produces a vector $p_R$ with entries in $\{0,1/t\}$ and $g(p_R)$ no smaller than before.

Let $R_t$ be the set obtained for cardinality $t$, and let $R_t^*$ be the minimum-cost $t$-set.
Since $p_{R_t^*}\in\Delta_t$, the preceding paragraphs give
\[
        g(p_{R_t})\ge g(p_{R_t^*})-\eta/4 .
\]
Equivalently,
\[
        F(R_t)
        \le F(R_t^*)+\frac{\eta}{4}Bt^2
        \le F(R_t^*)+\eta BH^2 .
\]
Enumerating $t=0,1,\ldots,H$ adds only a factor $n$ to the running time.
\end{proof}

\subsection{The QPTAS algorithm}

The QPTAS has three cases.
If the optimum is very small, Theorem~\ref{thm:smallopt} finds an optimum assignment exactly.
If the optimum is large, a standard additive approximation for dense CSPs already gives small relative error.
In the remaining intermediate range, the close-list theorem supplies a nearby reference assignment and Lemma~\ref{lem:quad} optimizes the quadratic part around it.

\begin{theorem}[Dense additive approximation~\cite{YZ14}]
\label{thm:dense-additive}
For every fixed arity $k$ and finite alphabet $\Sigma$, there is a deterministic algorithm that, given a complete $k$-CSP instance $\cJ$ over $\Sigma$ on $n$ variables and an accuracy parameter $\zeta\in(0,1)$, outputs an assignment of cost at most
\[
        \opt(\cJ)+\zeta\binom nk
\]
and runs in time $n^{\operatorname{poly}(1/\zeta)}$.
\end{theorem}

The next theorem combines the additive high-optimum case with the close-list and quadratic-search low-optimum case.

\begin{theorem}\label{thm:qptas}
If $1\le w(A)\le2$, then for every $\varepsilon\in(0,1)$ there is a deterministic algorithm running in time $n^{\operatorname{poly}(\log n,1/\varepsilon)}$ and returning $\widehat x$ with
\[
        \unsat_{\cI}(\widehat x)\le (1+\varepsilon)\opt(\cI).
\]
\end{theorem}

\begin{proof}
We may assume that $n$ is larger than a constant depending only on $k$ and $A$; otherwise exhaustive search takes constant time.
Since $w(A)\le2$, the parameter $r$ in Theorems~\ref{thm:pivot} and~\ref{thm:smallopt} is $2$, and hence $q=k-2$.
Let $\tau$ be the threshold from Theorem~\ref{thm:smallopt}.

Choose a sufficiently small constant $c>0$, depending only on $k$ and $A$, and set
\[
        T_{\mathrm{dense}}:=\frac{c\varepsilon^2 n^k}{\log^6 n},
        \qquad
        \zeta:=\frac{c\varepsilon^3}{\log^6 n},
        \qquad
        \eta:=\frac{c\varepsilon}{\log^2 n}.
\]
We choose $c$ small enough for the estimates below to hold.

We will use the following consequences of Lemma~\ref{lem:polyexpansion} and Theorem~\ref{thm:pivot}.
For every reference assignment $z$, the pair coefficients in Lemma~\ref{lem:polyexpansion} satisfy
\[
        \max_{\{i,j\}\in\binom V2}|a_{\{i,j\}}|=O(n^{k-2}),
\]
and, whenever $|R|\le H\le n$,
\[
        |E_z(R)|\le O(n^{k-3}H^3).
\]
Also, for every $T\ge \tau n^{k-2}$, Theorem~\ref{thm:pivot} gives
\[
        H_T=O\left(\left(\frac{T}{n^{k-2}}\right)^{1/2}\log n\right).
\]

\paragraph{Algorithm.}
Maintain a set $\mathcal C$ of candidate assignments.
First run Theorem~\ref{thm:smallopt}; if it returns an assignment, add it to $\mathcal C$.
Next run Theorem~\ref{thm:dense-additive} with accuracy $\zeta$, and add its output to $\mathcal C$.

Now consider the numbers
\[
        T_j=\min\left\{(1+\varepsilon/100)^j,\binom nk\right\},
        \qquad
        j=0,1,\ldots,\left\lceil \log_{1+\varepsilon/100}\binom nk\right\rceil .
\]
For each such $T_j$, construct the deterministic list $Z_{T_j}$ from Theorem~\ref{thm:pivot}(a).
For each $z\in Z_{T_j}$, compute the quadratic part $Q_z$ from Lemma~\ref{lem:polyexpansion}, and put
\[
        B_z:=\max\left\{1,\max_{\{i,j\}\in\binom V2}|a_{\{i,j\}}|\right\}.
\]
Apply Lemma~\ref{lem:quad} to $Q_z$ with $B=B_z$, $H=H_{T_j}$, and accuracy $\eta$.
Let $R\subseteq V$ be the set returned by the lemma, and add $z\oplus R$ to $\mathcal C$.
Finally, return an assignment in $\mathcal C$ minimizing $\unsat_{\cI}$.

\paragraph{Analysis.}
Let $x^*$ be an optimum assignment.
If $\opt(\cI)<\tau n^{k-2}$, then Theorem~\ref{thm:smallopt} returns an optimum assignment, so the algorithm succeeds.
We may therefore assume
\[
        \opt(\cI)\ge \tau n^{k-2}.
\]

The dense additive output has cost at most $\opt(\cI)+\zeta\binom nk$.
If $\opt(\cI)>T_{\mathrm{dense}}$, then
\[
        \opt(\cI)+\zeta\binom nk
        \le \opt(\cI)+\frac{c\varepsilon^3 n^k}{\log^6 n}
        = \opt(\cI)+\varepsilon T_{\mathrm{dense}}
        \le (1+\varepsilon)\opt(\cI).
\]
Thus it remains to consider the case
\[
        \tau n^{k-2}\le \opt(\cI)\le T_{\mathrm{dense}}.
\]

Choose $T=T_j$ from the above list so that
\[
        \opt(\cI)\le T\le (1+\varepsilon/100)\opt(\cI).
\]
Then $T\le 2T_{\mathrm{dense}}$, since $\varepsilon<1$.
By Theorem~\ref{thm:pivot}(a), the list $Z_T$ contains an assignment $z$ satisfying
\[
        \dist_H(z,x^*)\le H_T.
\]
Let
\[
        R^*:=\{i\in V:z_i\ne x^*_i\}.
\]
Then $|R^*|\le H_T$ and $z\oplus R^*=x^*$.

For this assignment $z$, Lemma~\ref{lem:quad} returns a set $R$ with $|R|\le H_T$ and
\[
        Q_z(R)
        \le Q_z(R^*)+\eta B_zH_T^2.
\]
By the coefficient bound and the bound on $H_T$,
\[
        B_zH_T^2
        =O(n^{k-2})\cdot
          O\left(\frac{T}{n^{k-2}}\log^2 n\right)
        =O(T\log^2 n).
\]
Hence
\[
        \eta B_zH_T^2
        =O(c\varepsilon T).
\]
By choosing $c$ small enough, this is at most $(\varepsilon/100)T$.

We next bound the higher-degree remainder.
Using the bound on $H_T$ again,
\[
\begin{aligned}
        n^{k-3}H_T^3
        &=
        O\left(
        n^{k-3}
        \left(\frac{T}{n^{k-2}}\right)^{3/2}
        \log^3 n
        \right)  \\
        &=
        O\left(
        T\left(\frac{T}{n^k}\right)^{1/2}\log^3 n
        \right).
\end{aligned}
\]
Since $T\le 2T_{\mathrm{dense}}=2c\varepsilon^2 n^k/\log^6 n$, this is
\[
        O(\sqrt c\,\varepsilon T).
\]
By choosing $c$ small enough, Lemma~\ref{lem:polyexpansion} gives
\[
        |E_z(R)|\le \frac{\varepsilon}{100}T,
        \qquad
        |E_z(R^*)|\le \frac{\varepsilon}{100}T.
\]

Finally, Lemma~\ref{lem:polyexpansion} gives
\[
        \unsat_{\cI}(z\oplus R)
        =
        \unsat_{\cI}(z)+Q_z(R)+E_z(R).
\]
Therefore
\[
\begin{aligned}
        \unsat_{\cI}(z\oplus R)
        &\le
        \unsat_{\cI}(z)+Q_z(R^*)
        +\eta B_zH_T^2+|E_z(R)|  \\
        &=
        \opt(\cI)-E_z(R^*)
        +\eta B_zH_T^2+|E_z(R)|  \\
        &\le
        \opt(\cI)+\frac{3\varepsilon}{100}T  \\
        &\le
        \opt(\cI)+\frac{3\varepsilon}{100}
        \left(1+\frac{\varepsilon}{100}\right)\opt(\cI)
        \le (1+\varepsilon)\opt(\cI).
\end{aligned}
\]
Thus $\mathcal C$ contains a $(1+\varepsilon)$-approximate assignment, and the algorithm returns one.

It remains only to check the running time.
The number of values $T_j$ is $O(\varepsilon^{-1}\log n)$.
For each such value, Theorem~\ref{thm:pivot}(a) constructs a list of size $n^{O(\log n)}$.
Lemma~\ref{lem:quad} is called with
\[
        \eta=\Theta(\varepsilon/\log^2 n),
\]
so each call takes time
\[
        n^{O(\log n/\eta^2)}
        =
        n^{\operatorname{poly}(\log n,1/\varepsilon)}.
\]
The dense additive algorithm is also quasi-polynomial, because
\[
        \zeta^{-1}
        =
        \frac{\log^6 n}{c\varepsilon^3}
        =
        \operatorname{poly}(\log n,1/\varepsilon).
\]
Hence the total running time is
\[
        n^{\operatorname{poly}(\log n,1/\varepsilon)}.
\]
\end{proof}

\section{Hardness}\label{sec:lower}
In this section, we prove the hardness part of Theorem~\ref{thm:main}.
Specifically, we show the $n^{w(A)-3-\eps}$-hardness for $w(A) \geq 4$ and $\omega(1)$-hardness for $w(A) = 3$ assuming the Unique Games Conjecture (UGC).

The reductions below have the same high-level structure.
We start from a hard source instance, keep its variables, and add one large auxiliary bucket.
For each source constraint, we add many $k$-ary constraints using the variables of that source constraint and variables from the bucket.
The literals are chosen so that a satisfied source constraint makes these $k$-ary constraints satisfied when all bucket variables are set to zero.
Because the output instance must be complete, every other $k$-set also receives a constraint.
The extra constraints with few source variables are chosen to be satisfied when all non-source variables in the $k$-set are set to zero, while the extra constraints with many source variables are few enough to be counted directly.
The soundness proofs show that a violated source constraint creates many violated constraints using the bucket.

The next lemma contains simple facts about $A$ used in the reductions.
It says that near an interval of accepted counts of maximum length, there is a rejected count.
This rejected count is used to choose literals in the constraints corresponding to source constraints.

\begin{lemma}\label{lem:boundary-counts}
Let $A$ be nontrivial.
\begin{enumerate}
\renewcommand{\labelenumi}{(\alph{enumi})}
\item If $w(A)\ge4$, then there is $g\in\{0,\ldots,k-3\}$ such that either
\[
        g\notin A,
        \qquad g+1,g+2,g+3\in A,
\]
or
\[
        g,g+1,g+2\in A,
        \qquad g+3\notin A.
\]

\item If $w(A)=3$, then there is $g\in\{0,\ldots,k-2\}$ such that either
\[
        g\notin A,
        \qquad g+1,g+2\in A,
\]
or
\[
        g,g+1\in A,
        \qquad g+2\notin A.
\]

\item If $A$ is complement-invariant, i.e. $t\in A$ if and only if $k-t\in A$, and $w(A)\ge3$, then an interval $[a,a+w(A)-1]\subseteq A$ of length $w(A)$ can be chosen with $a>0$. Consequently, $f=a-1$ satisfies $f\le k-w(A)$ as well as 
\[
        f,k-f\notin A,
        \qquad f+1,f+2, \dots, f+w(A) \in A.
\]
\end{enumerate}
\end{lemma}
\begin{proof}
Let $[a,a+w(A)-1]\subseteq A$ be an interval of length $w(A)$.
If $a>0$, maximality of $w(A)$ gives $a-1\notin A$.
For part (a), take $g=a-1$.
Since $w(A)\ge4$, the next three weights lie in the interval and $g\le k-w(A)\le k-4$.
For part (b), the same choice gives the first alternative.

The remaining case in parts (a) and (b) is $a=0$.
If $w(A)\ge4$, nontriviality gives $w(A)\le k$ and $w(A)\notin A$.
Taking $g=w(A)-3$ gives $g,g+1,g+2\in A$ and $g+3\notin A$.
If $w(A)=3$, then $k\ge3$ and $3\notin A$.
Taking $g=1$ gives $g,g+1\in A$ and $g+2\notin A$.

For part (c), choose an interval $[a,a+w(A)-1]\subseteq A$ of length $w(A)$.
If $a=0$, complement-invariance gives its reflection $[k-w(A)+1,k]$, also of length $w(A)$.
Since $A$ is nontrivial, $w(A)\le k$, so this reflected interval has positive left endpoint; replace the original interval by this reflected interval.
Now $a>0$. The claims on $f=a-1$ are immediate. 
\end{proof}

The next lemma explains how to fill unused $k$-sets that contain few source variables.
It guarantees satisfaction for every assignment to the source variables, as long as all other variables in the $k$-set are set to zero.

\begin{lemma}[Completion with few source variables]\label{lem:completion}
Let $[a,a+w(A)-1]\subseteq A$ be an interval of length $w(A)$.
Suppose a $k$-set contains $t\le w(A)-1$ source variables and $k-t$ other variables.
Then one can choose one literal on each variable in the $k$-set so that, whenever the other variables are all zero, the constraint is satisfied for every assignment to the $t$ source variables.
\end{lemma}

\begin{proof}
Since $a+w(A)-1\le k$ and $t\le w(A)-1$, we have $a\le k-t$.
Choose $a$ of the other variables to appear as literals of the form $1-x_i$, and choose the remaining other variables to appear as literals of the form $x_i$.
Then exactly $a$ of the other-variable literals are true when all other variables are zero.
The source variables can contribute any number between $0$ and $t$ true literals.
Thus the total number of true literals lies in
\[
        \{a,a+1,\ldots,a+t\}\subseteq \{a,a+1,\ldots,a+w(A)-1\}\subseteq A,
\]
so the constraint is satisfied.
\end{proof}

\subsection[Width at least four]{Width at least four}

We first prove the polynomial-factor hardness for predicates with width at least four.
The proof has two cases, depending on whether $A$ is invariant under complementing all literals.

\begin{lemma}[Hardness for width at least four]\label{lem:lower-poly}
Let $A$ be nontrivial with $w(A)\ge4$.
For every fixed $\varepsilon\in(0,w(A)-3)$, complete Min-CSP($A$) is NP-hard to approximate within factor $n^{w(A)-3-\varepsilon}$.
\end{lemma}

\begin{proof}
Choose an integer $K\ge 2w(A)/\varepsilon$.
The parameter $K$ is a constant because $k$, $A$, and $\varepsilon$ are fixed.
In both cases below, we use a standard source problem in which every constraint has three distinct variables and no two constraints use the same three-variable set.
This condition ensures that two different source constraints never try to define two constraints on the same output $k$-set.

\paragraph{Construction when $A$ is not complement-invariant.}
We reduce from the NP-hard problem of deciding satisfiability of such a 3-CNF instance.
A source constraint is a disjunction $\ell_1\vee\ell_2\vee\ell_3$, where each $\ell_i$ is either a source variable or its negation.
Under an assignment to the source variables, each $\ell_i$ has a value in $\{0,1\}$, and the constraint is violated exactly when the values of $\ell_1,\ell_2,\ell_3$ are all zero.
Fix an arbitrary order on the source variables, and list the three variables of each source constraint in that order.
Let $X$ be the source-variable set and let $m:=|X|$.
Set
\[
        Q:=2m^K .
\]
Choose $p\in A$ with $k-p\notin A$, which exists because $A$ is not complement-invariant.
The output variables are the variables in $X$ and one auxiliary bucket $Y$ of size $Q$.
We now specify the $A$-constraint on every $k$-set.

\begin{itemize}
\item Let $\mathfrak c$ be a source constraint, and let $\tau(\mathfrak c)\in\{0,1\}^3$ be the unique local assignment to its ordered variables that violates $\mathfrak c$.
For every $D\in\binom{Y}{k-3}$, add an $A$-constraint on the three variables of $\mathfrak c$ together with $D$.
Let $g$ be given by Lemma~\ref{lem:boundary-counts}(a).
Since $0\le g\le k-3$, choose literals on the variables in $D$ so that exactly $g$ literals from $D$ are true when all variables in $D$ are zero.
If $g\notin A$ and $g+1,g+2,g+3\in A$, choose on each source variable the literal that is false on its value in $\tau(\mathfrak c)$.
Then the number of true source literals is zero on $\tau(\mathfrak c)$ and is positive on every other local assignment.
If $g,g+1,g+2\in A$ and $g+3\notin A$, choose on each source variable the literal that is true on its value in $\tau(\mathfrak c)$.
Then the number of true source literals is three on $\tau(\mathfrak c)$ and is at most two on every other local assignment.
In either case, when all variables in $D$ are zero, this constraint is violated exactly when $\mathfrak c$ is violated.

\item For every $k$-set $S\subseteq Y$, add a constraint on $S$ whose number of true literals is $p$ when all variables in $S$ are zero and $k-p$ when all variables in $S$ are one.
This is obtained by using $p$ literals of the form $1-x_i$ and $k-p$ literals of the form $x_i$.
Since $p\in A$ and $k-p\notin A$, the all-zero assignment on $S$ satisfies the constraint, while the all-one assignment violates it.

\item Every remaining $k$-set receives one $A$-constraint.
If such a $k$-set contains at most $w(A)-1$ source variables, choose its literals using Lemma~\ref{lem:completion}, with the non-source variables in the $k$-set playing the role of the other variables in that lemma.
If the remaining $k$-set contains at least $w(A)$ source variables, choose its literals arbitrarily.
\end{itemize}
The first item assigns at most one constraint to each $k$-set because no two source constraints use the same three-variable set.
The second item uses only $k$-sets contained in $Y$, and the last item fills every unused $k$-set.
Thus the output is a complete Min-CSP($A$) instance with $n=m+Q=\Theta(Q)$ variables.

\paragraph{Construction when $A$ is complement-invariant.}
We reduce from the NP-hard problem of deciding satisfiability of NAE-3-SAT instances.
A source constraint is written as $\operatorname{NAE}(\ell_1,\ell_2,\ell_3)$, where each $\ell_i$ is either a source variable or its negation.
It is violated exactly when the values of $\ell_1,\ell_2,\ell_3$ are all equal.
Fix an arbitrary order on the source variables, and list the three variables of each source constraint in that order.
After this ordering, a local assignment to the three variables is a vector $\tau\in\{0,1\}^3$.
If $\tau$ violates the NAE constraint, then the bitwise complement $1-\tau$ also violates it.
We call the unordered set $\{\tau,1-\tau\}$ the complementary pair of the constraint.
There are four complementary pairs in total.
Let $X$ be the source-variable set, let $m:=|X|$, and set
\[
        Q:=2m^K .
\]
By Lemma~\ref{lem:boundary-counts}(c), choose $f$ such that
\[
        f\notin A,
        \qquad f+1,f+2\in A,
        \qquad k-f\notin A.
\]
Here $k-3\ge2$.
If $k-3=1$, complement-invariance and $w(A)\ge4$ would force all five weights $0,1,2,3,4$ to be accepted, contradicting nontriviality.

The output variables are the variables in $X$ and one auxiliary bucket $Y$ of size $Q$.
Partition $Y$ into two equal parts $Y^0\cup Y^1$; this is possible because $Q$ is even.
We now specify the $A$-constraint on every $k$-set.

\begin{itemize}
\item Let $\mathfrak c=\operatorname{NAE}(\ell_1,\ell_2,\ell_3)$ be a source constraint.
For every $D\in\binom{Y}{k-3}$, add an $A$-constraint on the three source variables of $\mathfrak c$ together with $D$.
Since $f\le k-4\le k-3$, choose literals on the variables in $D$ so that exactly $f$ auxiliary literals are true when all variables in $D$ are zero.
If $D$ is contained in one of the two parts $Y^0,Y^1$, use the source literals $\ell_1,\ell_2,\ell_3$.
If $D$ meets both parts, use the complemented source literals $1-\ell_1,1-\ell_2,1-\ell_3$.

\item Every remaining $k$-set receives one $A$-constraint.
If such a $k$-set contains at most $w(A)-1$ source variables, choose its literals using Lemma~\ref{lem:completion}, with the non-source variables in the $k$-set playing the role of the other variables in that lemma.
If the remaining $k$-set contains at least $w(A)$ source variables, choose its literals arbitrarily.
\end{itemize}
Again, the first item assigns at most one constraint to each $k$-set because no two source constraints use the same three-variable set, and the last item fills every unused $k$-set.
Thus the output has $n=m+Q=\Theta(Q)$ variables.

\paragraph{Completeness.}
Suppose the source instance is satisfiable.
Use a satisfying assignment on the source variables and set every auxiliary variable to zero.
In the non-complement-invariant construction, each constraint from the first item is satisfied because the corresponding source constraint is satisfied, and every constraint from the second item is satisfied because its all-zero count is $p\in A$.
In the complement-invariant construction, each source NAE constraint has either one or two true literals after applying the source literals.
The same is true after complementing all three source literals.
Since all auxiliary variables are zero, every constraint from the first item has total count $f+1$ or $f+2$, both in $A$.
In both constructions, every remaining constraint with at most $w(A)-1$ source variables is satisfied by Lemma~\ref{lem:completion}.
Therefore only remaining constraints whose $k$-sets contain at least $w(A)$ source variables may be violated.
There are at most
\[
        O\left(\sum_{t=w(A)}^k m^t Q^{k-t}\right)
        \le O(m^{w(A)} Q^{k-w(A)})
\]
such constraints, because $Q\ge m$ and $k$ is fixed.
Thus, when the source instance is satisfiable, the constructed instance has optimum at most $O(m^{w(A)}Q^{k-w(A)})$.

\paragraph{Soundness.}
Suppose the source instance is not satisfiable, and fix an arbitrary assignment to the output variables.

First consider the non-complement-invariant construction.
The restriction of the output assignment to $X$ violates some source constraint $\mathfrak c$.
If at least $Q/2$ variables in $Y$ are zero, then every $(k-3)$-set of zero variables in $Y$ gives a violated constraint from the first item for $\mathfrak c$.
This gives at least $\binom{\lceil Q/2\rceil}{k-3}=\Omega(Q^{k-3})$ violations.
Otherwise, more than $Q/2$ variables in $Y$ are one, and every $k$-set of such variables violates its constraint from the second item.
This gives $\Omega(Q^k)$ violations.
Hence, when the source instance is unsatisfiable, every output assignment has cost at least $\Omega(Q^{k-3})$.

Now consider the complement-invariant construction.
The restriction of the output assignment to $X$ violates some source constraint $\mathfrak c=\operatorname{NAE}(\ell_1,\ell_2,\ell_3)$.
For this constraint, the values of $\ell_1,\ell_2,\ell_3$ are all equal.
Let $U$ be the set of variables in $Y$ assigned this common value.
We count violated constraints from the first item associated with this source constraint.
If $|U\cap Y^0|\ge Q/4$ or $|U\cap Y^1|\ge Q/4$, then there are $\Omega(Q^{k-3})$ sets $D\subseteq U$ of size $k-3$ that are contained in one of the two parts.
For each such $D$, the constraint uses $\ell_1,\ell_2,\ell_3$.
If $\ell_1=\ell_2=\ell_3=0$, the source literals contribute $0$ and the auxiliary literals contribute $f$, so the total count is $f$.
If $\ell_1=\ell_2=\ell_3=1$, the source literals contribute $3$ and the auxiliary literals contribute $(k-3)-f$, so the total count is $k-f$.
Both counts are outside $A$, so all these constraints are violated.
Otherwise, both $Y^0\setminus U$ and $Y^1\setminus U$ have size larger than $Q/4$.
Since $k-3\ge2$, there are $\Omega(Q^{k-3})$ sets $D\subseteq Y\setminus U$ of size $k-3$ that meet both parts.
For each such $D$, the constraint uses $1-\ell_1,1-\ell_2,1-\ell_3$.
If $\ell_1=\ell_2=\ell_3=0$, the complemented source literals contribute $3$ and the auxiliary literals contribute $(k-3)-f$, so the total count is $k-f$.
If $\ell_1=\ell_2=\ell_3=1$, the complemented source literals contribute $0$ and the auxiliary literals contribute $f$, so the total count is $f$.
Again both counts are outside $A$, so all these constraints are violated.
Thus, when the source instance is unsatisfiable, every output assignment again has cost at least $\Omega(Q^{k-3})$.

Every unsatisfiable source instance therefore produces an instance with optimum at least $\Omega(Q^{k-3})$, whereas every satisfiable source instance produces one with optimum at most $O(m^{w(A)}Q^{k-w(A)})$.
The gap is therefore
\[
        \Omega\left(\frac{Q^{k-3}}{m^{w(A)}Q^{k-w(A)}}\right)
        =\Omega\left(\frac{Q^{w(A)-3}}{m^{w(A)}}\right).
\]
As $n=\Theta(Q)$ and $Q=2m^K$, this is
\[
        \Omega\left(n^{w(A)-3-w(A)/K}\right).
\]
The choice of $K$ gives $w(A)/K\le\varepsilon/2$.
For sufficiently large source instances, the hidden constant is absorbed, and the gap is at least $n^{w(A)-3-\varepsilon}$.
Thus an approximation algorithm with factor $n^{w(A)-3-\varepsilon}$ would distinguish satisfiable from unsatisfiable source instances, which is NP-hard.
\end{proof}

\subsection[Width-three]{Width-three}

For width-three $A$, we prove hardness by reducing from general Min-2-SAT instances.
The following source gap is used as a black box; we state it in the form where no two constraints use the same unordered pair of variables.


\begin{theorem}[UG-hardness of Min-2-SAT gap~\cite{Khot02}]\label{thm:binary-source-ugc}
Assuming the Unique Games Conjecture, for every constant $R>1$ there are constants $0<\beta<\gamma<1$ with $\gamma/\beta>R$ such that it is NP-hard to distinguish Min-2-SAT instances $\cJ$ with $M$ binary constraints, with no two constraints on the same unordered pair of variables, satisfying
\[
        \opt(\cJ)\le \beta M
        \qquad\text{from those satisfying}\qquad
        \opt(\cJ)\ge \gamma M.
\]
\end{theorem}

The next lemma gives the reduction from this Min-2-SAT gap to complete Min-CSP($A$) when $w(A)=3$.

\begin{lemma}\label{lem:binary-transfer-h3}
Let $A$ be nontrivial and satisfy $w(A)=3$.
Let $0<\beta<\gamma<1$ and let $\rho<\gamma/\beta$.
If it is NP-hard to distinguish Min-2-SAT instances $\cJ$ with $M$ binary constraints, no two on the same unordered pair of variables, and $\opt(\cJ)\le\beta M$ from those with $\opt(\cJ)\ge\gamma M$, then complete Min-CSP($A$) is NP-hard to approximate within factor $\rho$.
\end{lemma}

\begin{proof}
Let the source instance have variable set $X$, with $|X|=m$, and constraint set $E$, with $|E|=M$.
We may delete isolated source variables, so $m\le2M$.
Each source constraint $e\in E$ is a binary Min-2-SAT constraint on two distinct variables $u,v$; let $\tau^e\in\{0,1\}^{\{u,v\}}$ be its unique violating local assignment.
Set
\[
        Q:=(m+M+1)^{k+2}.
\]
Choose constants $\delta,\eta>0$ so that
\[
        \frac{\gamma(1-\delta)^{k-2}}{(1+\eta)\beta}>\rho .
\]

\paragraph{Construction when $A$ is not complement-invariant.}
Choose $p\in A$ with $k-p\notin A$, which exists because $A$ is not complement-invariant.
The output variables are the variables in $X$ and one auxiliary bucket $Y$ of size $Q$.
We specify the $A$-constraint on every $k$-set as follows.

\begin{itemize}
\item For each source constraint $e$ on variables $u,v$, and for every $D\in\binom{Y}{k-2}$, add an $A$-constraint on $\{u,v\}\cup D$.
Let $g$ be given by Lemma~\ref{lem:boundary-counts}(b).
Since $0\le g\le k-2$, choose literals on the variables in $D$ so that exactly $g$ literals from $D$ are true when all variables in $D$ are zero.
If $g\notin A$ and $g+1,g+2\in A$, choose on $u$ and $v$ the literals that are false on the values $\tau^e(u)$ and $\tau^e(v)$, respectively.
If $g,g+1\in A$ and $g+2\notin A$, choose on $u$ and $v$ the literals that are true on the values $\tau^e(u)$ and $\tau^e(v)$, respectively.
In either case, when all variables in $D$ are zero, this constraint is violated exactly when $x_u=\tau^e(u)$ and $x_v=\tau^e(v)$.

\item For each $k$-set $S\subseteq Y$, add a constraint on $S$ using $p$ literals of the form $1-x_i$ and $k-p$ literals of the form $x_i$.
Then the number of true literals is $p$ when all variables in $S$ are zero and $k-p$ when all variables in $S$ are one.
Thus the all-zero assignment on $S$ satisfies the constraint, while the all-one assignment violates it.

\item Every remaining $k$-set receives one $A$-constraint.
If such a $k$-set contains at most two source variables, choose its literals using Lemma~\ref{lem:completion}, with the non-source variables in the $k$-set playing the role of the other variables in that lemma.
If the remaining $k$-set contains at least three source variables, choose its literals arbitrarily.
\end{itemize}
Because no two source constraints use the same unordered pair, the first item never assigns two constraints to the same $k$-set.
The second item uses only $k$-sets contained in $Y$, and the last item fills every unused $k$-set.

\paragraph{Construction when $A$ is complement-invariant.}
By Lemma~\ref{lem:boundary-counts}(c), choose $f$ such that
\[
        f\notin A,
        \qquad f+1,f+2\in A,
        \qquad k-f\notin A.
\]
Here $k-2\ge2$; otherwise $k=3$, and complement-invariance together with $w(A)=3$ would make $A$ trivial.
The output variables are the variables in $X$, one auxiliary variable $r$, and one auxiliary bucket $Y$ of size $Q$.
We specify the $A$-constraint on every $k$-set as follows.

\begin{itemize}
\item For each source constraint $e$ on variables $u,v$, and for every $D\in\binom{Y}{k-2}$, add an $A$-constraint on $\{u,v\}\cup D$.
For each $i\in\{u,v\}$, use the literal $x_i$ if $\tau^e(i)=0$, and use the literal $1-x_i$ if $\tau^e(i)=1$.
Thus the chosen literal on $i$ is true exactly when the output value $x_i$ is different from $\tau^e(i)$.
Since $f\le k-3\le k-2$, choose literals on the variables in $D$ so that exactly $f$ auxiliary literals are true when all variables in $D$ are zero.
These choices have the following property, which is used in the analysis.
Suppose all variables in $D$ have value $x_r$.
If $x_r=0$, the auxiliary variables contribute $f$, so the total count is $f$ exactly when $x_u=\tau^e(u)$ and $x_v=\tau^e(v)$, and is $f+1$ or $f+2$ otherwise.
If $x_r=1$, the auxiliary variables contribute $k-2-f$, so the total count is $k-f$ exactly when $x_u=1-\tau^e(u)$ and $x_v=1-\tau^e(v)$, and is $k-(f+1)$ or $k-(f+2)$ otherwise.
Since $f,k-f\notin A$ and $f+1,f+2,k-(f+1),k-(f+2)\in A$, this constraint is violated exactly when $x_i\oplus x_r=\tau^e(i)$ for both $i\in\{u,v\}$.

\item For each $(k-1)$-set $S\subseteq Y$, add a constraint on $S\cup\{r\}$.
Use the literal $1-x_r$ on the variable $r$, and choose exactly $f$ literals on $S$ to be true when their variables are zero.
If all variables in $S$ equal $x_r$, then the total count is either $f+1$ or $k-(f+1)$, so the constraint is satisfied.
If all variables in $S$ equal $1-x_r$, then the total count is either $f$ or $k-f$, so the constraint is violated.

\item Every remaining $k$-set receives one $A$-constraint.
If such a $k$-set contains at most two source variables, choose its literals using Lemma~\ref{lem:completion}, with the anchor $r$ and the bucket variables in the $k$-set playing the role of the other variables in that lemma.
If the remaining $k$-set contains at least three source variables, choose its literals arbitrarily.
\end{itemize}
Again, because no two source constraints use the same unordered pair, the first item never assigns two constraints to the same $k$-set.
The second item uses $k$-sets containing $r$ and no source variable, and the last item fills every unused $k$-set.

\paragraph{Completeness.}
Suppose the source instance has an assignment violating at most $\beta M$ constraints.
Use this assignment on $X$.
In the complement-invariant construction, set $x_r=0$.
In both constructions, set all bucket variables to zero.
Every satisfied source constraint makes all constraints from the first item associated with it satisfied, by construction.
Every violated source constraint can create at most $\binom Q{k-2}$ violated constraints from the first item.
The constraints from the second item of the construction are satisfied, and all remaining constraints with at most two source variables are satisfied by Lemma~\ref{lem:completion}.
The only other possible violations are remaining constraints whose $k$-sets contain at least three source variables.
The number of such $k$-sets is
\[
        O\left(\sum_{t=3}^k m^t(Q+1)^{k-t}\right)
        \le O\left(m^3(Q+1)^{k-3}\right)
        =o\left(M\binom Q{k-2}\right),
\]
because $m\le2M$ and $Q=(m+M+1)^{k+2}$.
For sufficiently large source instances, this error term is at most $\eta\beta M\binom Q{k-2}$.
Thus, if $\opt(\cJ)\le\beta M$, the constructed instance has optimum at most
\[
        (1+\eta)\beta M\binom Q{k-2} .
\]

\paragraph{Soundness.}
Fix an arbitrary assignment to the output variables.
The condition $\opt(\cJ)\ge\gamma M$ means that every assignment to the source variables violates at least $\gamma M$ source constraints.
Apply this to the values $x_i$ on the source variables in the non-complement-invariant construction, and to the values $x_i\oplus x_r$ in the complement-invariant construction.
Thus, in the non-complement-invariant construction, at least $\gamma M$ source constraints $e$ on variables $u,v$ satisfy
\[
        x_u=\tau^e(u)
        \qquad\text{and}\qquad
        x_v=\tau^e(v).
\]
In the complement-invariant construction, at least $\gamma M$ source constraints $e$ on variables $u,v$ satisfy
\[
        x_i\oplus x_r=\tau^e(i)
        \qquad\text{for both } i\in\{u,v\}.
\]

First consider the non-complement-invariant construction.
If at least $(1-\delta)Q$ variables in $Y$ are zero, then each violated source constraint $e$ on variables $u,v$ contributes a violated constraint for every $(k-2)$-set of zero variables in $Y$.
The $k$-sets counted for different violated source constraints are distinct because no two source constraints use the same unordered pair.
Hence there are at least
\[
        \gamma M \binom{\lceil(1-\delta)Q\rceil}{k-2}
        =(1-o(1))\gamma M(1-\delta)^{k-2}\binom Q{k-2}
\]
violations.
Otherwise, at least $\delta Q$ variables in $Y$ are one, and the constraints from the second item on $k$-sets of such variables give at least
\[
        \binom{\lceil\delta Q\rceil}{k}=\Omega(Q^k)=\omega\left(M\binom Q{k-2}\right)
\]
violations.

Now consider the complement-invariant construction.
If at least $(1-\delta)Q$ variables in $Y$ are equal to $x_r$, then each violated source constraint contributes a violated constraint for every $(k-2)$-set of such variables.
Again, the counted $k$-sets are distinct for different violated source constraints, so this gives at least
\[
        \gamma M \binom{\lceil(1-\delta)Q\rceil}{k-2}
        =(1-o(1))\gamma M(1-\delta)^{k-2}\binom Q{k-2}
\]
violations.
Otherwise, at least $\delta Q$ variables in $Y$ are equal to $1-x_r$, and the constraints from the second item on $(k-1)$-sets of such variables give at least
\[
        \binom{\lceil\delta Q\rceil}{k-1}=\Omega(Q^{k-1})=\omega\left(M\binom Q{k-2}\right)
\]
violations.

Therefore, if $\opt(\cJ)\ge\gamma M$, every output assignment has cost at least
\[
        (1-o(1))\gamma M(1-\delta)^{k-2}\binom Q{k-2} .
\]
For sufficiently large source instances, the ratio between the lower bound when $\opt(\cJ)\ge\gamma M$ and the upper bound when $\opt(\cJ)\le\beta M$ is larger than $\rho$, by the choice of $\delta$ and $\eta$.
Since $k$ is fixed and $Q$ is polynomial in the source size, the reduction is polynomial.
Thus a $\rho$-approximation for complete Min-CSP($A$) would distinguish instances with $\opt(\cJ)\le\beta M$ from instances with $\opt(\cJ)\ge\gamma M$.
\end{proof}

\end{document}